\documentclass[12pt,a4paper]{article}
\pdfoutput=1

\usepackage[utf8]{inputenc}
\usepackage[T1]{fontenc}

\usepackage{multirow}
\usepackage{cite}
\usepackage{amsmath}
\usepackage{color}
\usepackage{amsfonts}
\usepackage{amssymb}
\usepackage{graphicx}
\usepackage{geometry}
\usepackage{amssymb,epsfig,subfigure}
\usepackage{hyperref}
\usepackage{url}
\usepackage{comment}
\usepackage[font=footnotesize]{caption}
\usepackage{slashed}

\makeatletter
\renewcommand\section{\@startsection {section}{1}{\z@}%
                                 {-3.5ex \@plus -1ex \@minus -.2ex}
                                   {2.3ex \@plus.2ex}%
                                   {\normalfont\large\bfseries}}
\renewcommand\subsection{\@startsection{subsection}{2}{\z@}%
                                   {-3.25ex\@plus -1ex \@minus -.2ex}%
                                     {1.5ex \@plus .2ex}%
                                     {\normalfont\bfseries}}
\renewcommand\subsubsection{\@startsection{subsubsection}{3}{\z@}%
                                   {-3.25ex\@plus -1ex \@minus -.2ex}%
                                     {1.5ex \@plus .2ex}%
                                     {\normalfont\itshape}}
\makeatother

\def\pplogo{\vbox{\kern-\headheight\kern -29pt
\halign{##&##\hfil\cr&{\ppnumber}\cr\rule{0pt}{2.5ex}&\ppdate\cr}}}
\makeatletter
\def\ps@firstpage{\ps@empty \def\@oddhead{\hss\pplogo}%
  \let\@evenhead\@oddhead 
}
\def\maketitle{\par
 \begingroup
 \def\thefootnote{\fnsymbol{footnote}}
 \def\@makefnmark{\hbox{$^{\@thefnmark}$\hss}}
 \if@twocolumn
 \twocolumn[\@maketitle]
 \else \newpage
 \global\@topnum\z@ \@maketitle \fi\thispagestyle{firstpage}\@thanks
 \endgroup
 \setcounter{footnote}{0}
 \let\maketitle\relax
 \let\@maketitle\relax
 \gdef\@thanks{}\gdef\@author{}\gdef\@title{}\let\thanks\relax}
\makeatother

\numberwithin{equation}{section}

\newcommand\eea{\end{eqnarray}}
\newcommand\bea{\begin{eqnarray}}

\def\beq{\begin{equation}}
\def\eeq{\end{equation}}

\newcommand{\be}{\begin{equation}}
\newcommand{\ee}{\end{equation}}
\newcommand{\ba}{\begin{align}}
\newcommand{\ea}{\end{align}}
\newcommand{\bg}{\begin{gather}}
\newcommand{\eg}{\end{gather}}
\newcommand{\bseq}{\begin{subequations}}
\newcommand{\eseq}{\end{subequations}}

\newcommand{\coment}[1]{}

\begin{document}
\setcounter{page}0
\def\ppnumber{\vbox{\baselineskip14pt
}}
\def\ppdate{
} \date{}

\author{Lucas Daguerre$^1$\\
[7mm] \\
{\normalsize \it $^1$ Department of Physics and Astronomy, University of California, Davis}\\
{\normalsize \it Davis, California 95616, USA}}

\bigskip
\title{\bf Boundary correlators and the Schwarzian mode
\vskip 0.5cm}
\maketitle

\begin{abstract}
The effective low temperature dynamics  of near-extremal black holes is governed by the quantum fluctuations of the Schwarzian mode of JT gravity. Utilizing as a proxy a planar charged black hole in asymptotically Anti-de-Sitter spacetime, we investigate the effects of these fluctuations on a probe scalar field. The corresponding holographic real-time boundary correlators are computed following a holographic renormalization procedure, using the dubbed gravitational Schwinger-Keldysh geometry (grSK) and known exact results of boundary correlators from the near-horizon region. This analysis gives rise to a retarded Green's function that decays as a power law for late Lorentzian times. Its analytic structure indicates the presence of a branch cut in the complex frequency domain at finite temperature. These features are a non-perturbative hallmark that prevails as long as the planar transverse space is kept compact.
\end{abstract}

\bigskip\bigskip\bigskip\bigskip\bigskip\bigskip
{\small{\vspace{1.5 cm}\noindent ${}^{1}\,\,$ldaguerre@ucdavis.edu  }}

\newpage

\tableofcontents

\vskip 1cm

\section{Introduction}\label{sec:Intro}

Black holes are a robust prediction of General Relativity and provide a quantitative setting for studying quantum effects in semiclassical gravity. Of notable interest are near-extremal black holes, which can attain arbitrary low temperatures \cite{Turiaci:2023wrh}. Historically, several long standing puzzles have been addressed in relation to them: the breakdown of the statistical thermodynamic intepretation \cite{Preskill:1991tb}; the presence of a mass gap between an extremal black hole and the lightest near-extremal black hole \cite{Almheiri:2016fws}; a non-trivial entropy for extremal non-supersymmetric black holes, which is in tension with the third law of thermodynamics \cite{Page:2000dk}; and instabilities of extremal black holes under perturbations\cite{Aretakis:2012ei}. Recently, the thermodynamic puzzles have been revisited using an effective theory for the dynamics of the s-wave sector of near-extremal black holes known as JT gravity (Jackiw–Teitelboim) \cite{Jackiw:1984je,Teitelboim:1983ux}. For a review, refer to \cite{Mertens:2022irh}. This theory arises on the grounds that the generic near-horizon region of extremal black holes has topology AdS$_2\times \mathcal{R}$, where $\mathcal{R}$ is a compact manifold and AdS$_2$ is a 2-dimensional Anti-de-Sitter (AdS) spacetime. The emergent AdS$_2$ throat has enhanced $SL(2,\mathbb{R})$ symmetry, and the associated zero modes (henceforth Schwarzian mode) get an action by considering thermal deviations away from extremality \cite{Maldacena:2016upp}. The Schwarzian mode lives at the boundary of the near-horizon region, and its quantum fluctuations become strongly coupled at low temperatures \footnote{Indeed, the effective coupling in the JT gravity action is $C/\beta$\cite{Maldacena:2016upp}, where $\beta$ is the inverse of the black hole temperature and $C$ is a model dependent coupling (for instance, see (\ref{eq:scaleC}) for a charged planar black hole in asymptotically AdS). Precisely, the above statement in the text indicates that the Schwarzian mode becomes strongly coupled when $C/\beta \lesssim 1$.}. In fact, the euclidean black hole partition function of JT gravity turns out to be one-loop exact \cite{Stanford:2017thb}. As a result, black hole thermodynamic quantities, such as the energy and the entropy, get corrections at low temperatures. These corrections are present for both non-supersymmetric cases in \cite{Iliesiu:2020qvm,Iliesiu:2022onk,Banerjee:2023quv,Kapec:2023ruw,Rakic:2023vhv} as well as, for supersymmetric cases in \cite{Heydeman:2020hhw,Boruch:2022tno,Iliesiu:2022kny,Boruch:2023trc}. 

The AdS/CFT correspondence provides another framework that can be exploited to study quantum effects coming from near-extremal black holes in asymptotically Anti-de-Sitter spacetime. In a nutshell, the correspondence states that quantum gravity in asymptotically AdS$_{d+1}$ is dual to a strongly coupled $d$-dimensional field theory living at the conformal boundary of the spacetime 
\cite{Maldacena:1997re,Witten:1998qj,Gubser:1998bc}.
 Henceforth, our interest resides in computing holographic boundary correlators in a strongly coupled thermal field theory by doing computations with a probe scalar field in the dual semiclassical near-extremal geometry. The evaluation of euclidean boundary correlators has been among the early checks of the correspondece \cite{Aharony:1999ti}. Moreover, the first prescription regarding real-time correlators was given by \cite{Son:2002sd}, which employs ingoing boundary conditions on the future black hole horizon to calculate the holographic retarded Green's function. An overview of holographic real-time correlators can be found in the introduction of \cite{Jana:2020vyx}.


The novel contribution in this article is the inclusion of quantum effects on causal (real-time) holographic boundary correlators at low temperatures. So far, at the level of two-point functions, the near-horizon Wightman boundary correlators have already been computed at all orders in perturbation theory\cite{Mertens:2017mtv,Lam:2018pvp,Bagrets:2016cdf}.
In a straightforward manner, we were able to construct the associated exact retarded Green's function (\ref{eq:retardedJT}). This function has the property that decays as a power law for late Lorentzian times, unlike the characteristic exponential decay behavior dictated by the quasinormal modes\cite{Horowitz:1999jd,Birmingham:2001pj,Berti:2009kk}. In terms of the analytic structure in the complex frequency domain, the power law tail corresponds to the presence of a branch cut in the lower half-plane running from the origin. Further expressions for the retarded Green's function were obtained for the upper half-plane (\ref{eq:GR_upper_half}) and the real line (\ref{eq:GR_real_line}). The explicit close form expression for the lower half plane is still unknown. As a remedy, in Subsection \ref{subsub:ToyModel} we present a toy model for the retarded Green's function that captures the late Lorentzian time decay, whose close form Fourier transform can be analytically continued to the whole complex plane. The result indicates that the toy model contains a non-meromorphic function that is proportional to $\exp(-\frac{2\pi^2C}{\beta})$, signaling the non-perturbative nature of the quantum corrections at low temperature\cite{Mertens:2020pfe}, which also vanish in the semiclassical limit $C \to \infty$ for a fixed $\beta$.

Ultimately, the holographic boundary correlators of the $d$-dimensional field theory can be computed following the lines of the holographic renormalization procedure \cite{Heemskerk:2010hk,Faulkner:2010jy,Faulkner:2009wj}. This prescription uses the emergent radial direction of AdS to geometrize the energy scale of the dual field theory. As a consequence, we split the spacetime geometry in two parts: the near-horizon region (IR) and the far away region (UV).  The IR is dominated by the low temperature description, i.e. JT gravity, so that the boundary near-horizon retarded Green's function is again given by the exact result (\ref{eq:retardedJT}). On the other hand, the UV can be integrated out by solving a double-Dirichlet boundary condition problem for the probe scalar field, with physical sources at the asymptotic AdS$_{d+1}$ boundary and with transient sources at the boundary of the near-horizon region. The integration of the transient source yields a boundary retarded Green's function that is a rational function of the near-horizon retarded Green's function in momentum space. This prescription is, in spirit, semi-holographic \cite{Faulkner:2010tq,Faulkner:2010jy} because of the splitting of the degrees of freedom UV/IR in the bulk and the emergence of an effective theory in the IR. We opted to implement this matching procedure in real-time, using a complex geometry, the dubbed gravitational Schwinger-Keldysh geometry (grSK) introduced in \cite{Glorioso:2018mmw,deBoer:2018qqm} and further developed in \cite{Jana:2020vyx}, instead of other constructions such as \cite{Skenderis:2008dh,Skenderis:2008dg}.

The article is divided as follows. Section \ref{sec:Planar_RN} introduces the black-hole geometry. Subsection \ref{subsec:SetUp} describes features of the planar Reissner-Nordström spacetime in asymptotically AdS$_{d+1}$, including details about the near-extremal limit. Subsection \ref{subsec:effectiveJT} presents the effective two dimensional description of the spacetime at low temperatures in terms of the JT gravity action, as well as some recently understood implications regarding black hole thermodynamics. Section \ref{sec:BdyCorrelators} is completely devoted to holographic boundary correlators: Subsection \ref{subsec:correlatorsNHR} introduces the exact near-horizon region boundary correlators and the associated retarded Green's function. Subsection \ref{subsec:matchingUV} explains how to transfer the near-horizon correlators information to the asymptotic AdS$_{d+1}$ boundary. First, we provide a brief introduction to the Schwinger-Keldysh formalism in field theory in Subsection \ref{subsec:SKfieldtheory}. Then, we explain how to perform the calculation in the bulk using the gravitational Schwinger-Keldysh geometry. We show the result in a low momentum expansion for the  Reissner-Nordström-AdS$_{d+1}$ case in Subsection \ref{subsub:PlanarRN}, and at all orders for the three dimensional rotating BTZ black hole in Subsection \ref{subsub:RotatingBTZ}. Subsection \Ref{subsub:anal_structure} analyzes the analytic structure of the retarded Green's function in the complex frequency domain, and presents a toy model for the retarded Green's function with a close form Fourier transform in Subsection \ref{subsub:ToyModel}. Section \ref{sec:discussion} provides a conclusion and gives directions for future work.

In addition, the appendices are divided as follows. In Appendix \ref{app:planarRNapp} we provide complementary details regarding the Reissner-Nordström black-hole background. In Appendix \ref{subsec:geom_low_T} we elaborate further aspects of the geometry at low temperatures. In Appendix \ref{subsec:dim_reduction} a derivation of the effective JT gravity action is presented for the s-wave sector of the near-extremal black hole. In Appendix  \ref{app:BTZ} we show the solutions to the scalar wave equation for the rotating BTZ black hole. This include the explicit ingoing and outgoing solutions, as well as their asymptotics towards the near-horizon region and towards the AdS$_3$ boundary.
\section{Black hole geometry}
\label{sec:Planar_RN}
\subsection{Setup}
\label{subsec:SetUp}
The geometry of interest is given by a $(d+1)$-dimensional electrically charged black hole in asymptotically Anti-de-Sitter (AdS$_{d+1}$) spacetime, usually referred as the Reissner-Nordström-AdS$_{d+1}$ spacetime $\mathcal{M}_{d+1}$. The corresponding Einstein-Maxwell equations are obtained by varying the gravitational action that includes the Gibbons-Hawking boundary term evaluated at $\partial \mathcal{M}_{d+1}$ \cite{Chamblin:1999tk,Almheiri:2016fws}
\begin{equation}
\begin{split}
I_{\text{grav+matter}}&=-\frac{1}{16\pi G_N^{(d+1)}}\int_{\mathcal{M}_{d+1}} d^{d+1}x \sqrt{g^{(d+1)}}\left(R^{(d+1)}+\frac{d(d-1)}{\ell_{d+1}^2}-F_{AB}F^{AB}\right)\\
&\:\:\:\:\:\:\:\:-\frac{1}{8\pi G_N^{(d+1)}}\int_{\partial\mathcal{M}_{d+1}} d^{d}x \sqrt{\gamma^{(d)}}\:K^{(d)}\:,
\end{split}
\label{eq:EHM_action}
\end{equation}
where $\ell_{d+1}$ is the AdS$_{d+1}$ radius and $G_N^{(d+1)}$ is the $(d+1)$-dimensional Newton's constant. We follow the conventions for uppercase Latin indices $A,B=0,1,\dots,d$ and lowercase Latin indices $i,j=1,\dots,d-1$, where $d$ is the spacetime dimension of the boundary field theory. The resulting set of equations are
\begin{equation}
\begin{split}
&R_{AB}-\frac{1}{2}g_{AB}R-\frac{d(d-1)}{2\ell_{d+1}^2} g_{AB}=2\left(F_{AC}F_{BD}g^{CD}-\frac{1}{4}g_{AB}F_{CD}F^{CD}\right)\,,\\
&\nabla _A F^{AB}=0\,.
\end{split}
\label{eq:Einst_Maxw}
\end{equation}
A family of solutions to the Einstein-Maxwell equations (\ref{eq:Einst_Maxw}) are described by 
\begin{equation}
ds^2=f(r)d\tau^2+\frac{dr^2}{f(r)}+r^2\delta_{ij}dx^idx^j\,,\:\:\:\:\:\:\:\:\:\:\:\:\:\:\: \textbf{A}=A_{\tau}d\tau\,,
\label{eq:LineEl_Vect}
\end{equation}
with a redshift factor and Maxwell vector field
\begin{equation}
f(r)=\frac{r^2}{\ell_{d+1}^2}-\frac{m}{r^{d-2}}+\frac{q^2}{r^{2d-4}},\:\:\:\:\:\:\:\:\:\:\:\:A_{\tau}=-i \sqrt{\frac{d-1}{2(d-2)}}\frac{q}{r_+^{d-2}} \left(1-\frac{r_+^{d-2}}{r^{d-2}}\right)\,.
\label{eq:redshift_vectorP}
\end{equation}
The outer and inner horizons are denoted as $r_{\pm}$, respectively, and are zeros of $f(r_{\pm})=0$. Hence, the expressions (\ref{eq:LineEl_Vect}) and (\ref{eq:redshift_vectorP}) determine a 2-parameter family of solutions depending on $m$ and $q$ for a given AdS radius. To regulate infrared divergences, the transverse planar space is compactified in a torus $T^{d-1}$, such that the transverse coordinates are identified via $x^i \sim x^i +L$, for a dimensionless length $L$.
The mass $M$ and charge $Q$ of the black hole are
\begin{equation}
M=\frac{(d-1)}{16\pi G_N^{(d+1)}}m L^{d-1},\:\:\:\:\:\:\:\: Q=\frac{\sqrt{2(d-1)(d-2)}}{8\pi G_N^{(d+1)}}q L^{d-1}.
\end{equation}
Also, the Hawking temperature and the chemical potential are
\begin{equation}
T=\frac{d r_+^{2d-2}-(d-2)q^2 \ell_{d+1}^2}{4\pi \ell_{d+1}^2 r_{+}^{2d-3}}\,,\:\:\:\:\:\:\:\:\:\:\:\:\:\mu=\sqrt{\frac{d-1}{2(d-2)}}\frac{q}{r_+^{d-2}}\,.
\label{eq:temp_chem_pot}
\end{equation}
We are interested in studying the near-extremal (i.e. low temperature) description of the spacetime. For details refer to Appendix \ref{subsec:geom_low_T}, in this section we opted to summarize the main ingredients. At extremality the temperature is $T=0$ and the inner and outer horizons coincide at $r_0\equiv r_+=r_-$. The spacetime can then be divided in two parts: the near-horizon region satisfying $r-r_0 \ll r_0$, and the far away region satisfying $r_0 \ll r$. Notably, the geometry develops an infinite throat so that the topology of the near-horizon region becomes AdS$_2\times T^{d-1}$, with an effective AdS$_2$ length scale $\ell_2=\frac{\ell_{d+1}}{\sqrt{d(d-1)}}$. 
\begin{figure}[t]
    \centering
    \includegraphics[width=0.45\textwidth]{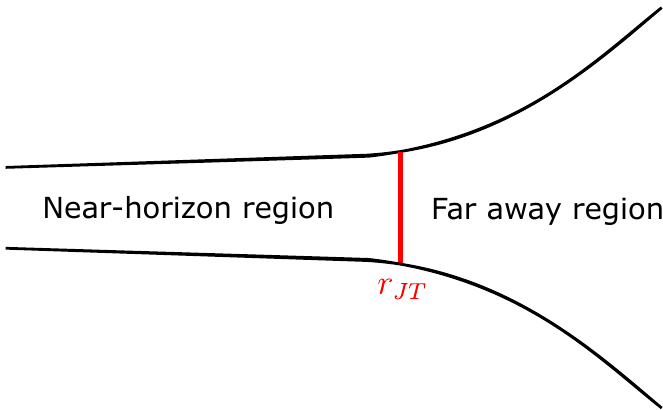}
    \caption{Diagram of a near-extremal black hole. At low temperatures, the geometry develops an infinite throat in the near-horizon region that slightly deviates from AdS$_2\times T^{d-1}$, whereas the far away region is well approximated by the extremal metric. For large black holes (with $\ell_{d+1} \ll r_0$), both regions overlaps at a hypersurface located at $r=r_{JT}$. The effective low temperature black hole dynamics is governed by the Schwarzian mode that lives at the boundary of the near-horizon region at $r=r_{JT}$.}
    \label{fig:sketchBH}
\end{figure}

The departure from extremality is achieved by increasing the temperature (or equivalently the mass) while keeping the charge $q(r_0)$ fixed (\ref{eq:Ext_mass_charge}), for a fixed value of $r_0$. A representation of the near-extremal geometry is depicted in Fig. \ref{fig:sketchBH}. The technical requirement to be close to extremality is given by (\ref{eq:Temp_cond}), which we reproduce here:
\begin{equation}
T\ell_2^2 \ll \ell_2 \ll r_0\:.
\label{eq:low_T_condition}
\end{equation}
Furthermore, the near-horizon region and the far away region are glued at a hypersurface with $r=r_{JT}$ such that  (\ref{eq:location_rJT}) holds, also being recast here \footnote{It is worth addressing the flat space limit where the cosmological constant vanishes, or equivalently, where the AdS$_{d+1}$ scale goes to $\ell_{d+1}\to \infty$. Such a limit does make sense for black holes with spherical horizons, as it was discussed for the 4-dimensional case in \cite{Iliesiu:2020qvm}, where the near-horizon geometry is $\text{AdS}_2 \times S^{2}$ and the effective $\text{AdS}_2$ length scale  is $\ell_2 \approx r_0$. However, the flat space limit does not make sense for black holes with planar horizons, such as those studied in this article, where the near-horizon geometry is $\text{AdS}_2 \times T^{2}$ and the effective $\text{AdS}_2$ length scale is $\ell_2=\frac{\ell_{4}}{\sqrt{6}}$. Specifically, $\ell_2$ is ill-defined if we take the limit $\ell_4 \to \infty$.}
\begin{equation}
 \ell_2 \ll r_{JT}-r_0\ll r_0\:.
\label{eq:location_rJT_sec2}
\end{equation}
In Section \ref{subsec:effectiveJT}, we will describe how an effective theory emerges by considering quantum fluctuations coming from the near-horizon throat. This theory is ruled by the Schwarzian mode living at a hypersurface with $r=r_{JT}$.
\subsection{Effective low temperature description}
\label{subsec:effectiveJT}
One way of capturing the effects of quantum fluctuations coming from the near-horizon region at low temperatures involves a dimensional reduction of the black hole geometry in the s-wave sector \cite{Iliesiu:2020qvm,Nayak:2018qej,Moitra:2019bub,Ghosh:2019rcj,Davison:2016ngz,Sachdev:2019bjn,Banerjee:2021vjy,Banerjee:2023quv}. In light of that, we propose the following ansatz that includes a two-dimensional metric $g_{\mu \nu}(x^{\rho})$ that depends on the coordinates $x^{\rho}$ with $\rho=(\tau,r)$, as well as a dilaton field $\Phi(x^{\rho})$ that measures the size of the transverse space, which is independent of the transverse coordinates $y^i$ with $i=1,\dots,d-1$. The line element is 
\begin{equation}
    ds^2=g_{AB}dx^Adx^B=g_{\mu \nu}(x^{\rho})dx^{\mu}dx^{\nu}+\Phi^2(x^{\rho}) \delta_{ij}dy^i dy^j\:,
    \label{eq:ansatz}
\end{equation}
and the Maxwell field is
\begin{equation}
A_{M}=(A_{\mu}(x^{\rho}),0)\:.
\label{eq:MaxwellSwave}
\end{equation}
After a careful but straightforward procedure, the effective on-shell action near extremality in the s-wave sector becomes (refer to Appendix \ref{subsec:dim_reduction} for a detailed derivation)
\begin{equation}
    I^{eff}_{\text{grav+matter}}=\beta M_0-S_0-C\int_0^{\beta} du \:\text{Sch}\left\{\tan\left(\frac{\pi \tau(u)}{\beta}\right),u\right\},
    \label{eq:actiondimRed}
\end{equation}
where $\text{Sch}\{f(u),u\}=\frac{f'''(u)}{f'(u)}-\frac{3}{2}\left(\frac{f''(u)}{f'(u)}\right)^2$ is the Schwarzian derivative of $f(u)$ with respect to the boundary time $u$, $\beta=T^{-1}$ is the inverse of the Hawking temperature, $\tau(u)$ is the boundary mode or Schwarzian mode, $M_0$ is the extremal mass, $S_0$ is the naive extremal entropy and $C=(2\pi^2 M_{\text{gap}})^{-1}$ is an effective scale related to the inverse of the "mass gap". The result (\ref{eq:actiondimRed}) is universal for a large class of near-extremal black holes \cite{Nayak:2018qej,Moitra:2019bub,Iliesiu:2020qvm,Iliesiu:2022onk,Banerjee:2023quv}. For the planar Reissner-Nordström-AdS$_{d+1}$ case, the constants are$\:$\footnote{In the case that the toroidal transverse space is decompactified taking $L\to \infty$, the effective coupling $C\to \infty$. At fixed temperature, this corresponds to the semiclassical limit where quantum effects of the Schwarzian mode are suppressed.}
\begin{equation}
  M_0=\frac{2(d-1)^2}{(d-2)}\frac{L^{d-1}}{16\pi G_N^{(d+1)}}\frac{r_0^d}{\ell_{d+1}^2},\:\:\:\:\:\:\:\:\:\:\:\:S_0=\frac{r_0^{d-1}L^{d-1}}{4 G_N^{(d+1)}}\:,\:\:\:\:\:\:\:\:C=\frac{1}{2\pi d}\frac{\ell_{d+1}^2 r_0^{d-2}L^{d-1}}{4G_N^{(d+1)}}\:.
  \label{eq:scaleC}
\end{equation}
The effective theory (\ref{eq:actiondimRed}), JT gravity (Jackiw–Teitelboim), gives an action for the Schwarzian mode $\tau(u)$ \cite{Maldacena:2016upp}, which is located at a hypersurface with $r=r_{JT}$ constrained by (\ref{eq:location_rJT_sec2}). The range of validity of (\ref{eq:actiondimRed})  requires that the temperature is bounded by $\mathcal{O}\left(r_0e^{-S_0}\right)\ll T \ll r_0$ in AdS units\cite{Iliesiu:2022onk}. The lower bound comes from other non-perturbative corrections that should be taken into account at that energy scale. Accordingly, this is the reason why $S_0$ is a naive extremal entropy, since we cannot trust the effective description down to $T=0$. On the other hand, using (\ref{eq:actiondimRed}) it is possible to compute the black hole partition function by employing the Gibbons-Hawking euclidean path integral \cite{Gibbons:1976ue}. In the present case, the partition function is 1-loop exact by considering quantum fluctuations of the Schwarzian mode \cite{Stanford:2017thb}. Hence the entropy and energy, in the canonical ensemble for a sector of fix charge, turns out to be 
\begin{equation}
    S=\underbrace{S_0+4\pi^2 C T}_{\text{classical}}+\underbrace{\frac{3}{2}\log\left(C T\right)+\frac{3}{2}}_{\text{quantum}}+\ldots \:,
    \label{eq:S_corrected}
\end{equation}
\begin{equation}
    E=\underbrace{M_0+2\pi^2 C  T^2}_{\text{classical}}+\underbrace{\frac{3}{2}T}_{\text{quantum}}+\ldots \:,
    \label{eq:E_corrected}
\end{equation}
where $C=(2\pi^2 M_{\text{gap}})^{-1}$ and the dots stand for further subleading polynomial contributions in temperature. The first part of (\ref{eq:S_corrected}) and (\ref{eq:E_corrected}) comes from the classical analysis of the euclidean path integral, whereas the second part comes from the 1-loop exact Schwarzian contribution to the partition function. There are other contributions coming from 1-loop matter field determinants \cite{Iliesiu:2022onk} that we are not displaying in the formulas (\ref{eq:S_corrected}) and (\ref{eq:E_corrected}). Furthermore, the inverse Laplace transform of (\ref{eq:E_corrected})  gives a density of states $\rho(E)$ that does not present a mass gap at $M_{\text{gap}}$ \cite{Iliesiu:2020qvm}. As noted in \cite{Iliesiu:2020qvm}, this clarifies the breakdown of the statistical mechanics interpretation of black holes at $T\sim M_{\text{gap}}$ \cite{Preskill:1991tb}, as the scale at which quantum fluctuations of the Schwarzian mode become strongly coupled.
\section{Boundary correlators}
\label{sec:BdyCorrelators}
As described in Section \ref{subsec:effectiveJT}, at low temperatures the dynamics of the s-wave sector of a Reissner-Nordström-AdS$_{d+1}$ black hole is governed by the quantum fluctuations of the Schwarzian mode. To probe these quantum effects using correlators, we introduce a scalar field $\varphi$ in the $(d+1)$-dimensional black hole background. According to the rules of the holographic AdS/CFT dictionary, the scalar field is dual to a primary operator $\mathcal{O}$ of scaling dimension $\Delta$ in the $d$-dimensional boundary field theory. The Gaussian action for the scalar field is given by
\begin{equation}
    I_{\text{scalar}}=\frac{1}{2}\int_{\mathcal{M}_{d+1}} d^{d+1}x \sqrt{g^{(d+1)}}\:\big(g^{AB}\partial_{A}\varphi\partial_{B}\varphi+m^2\varphi^2\big)\:,
    \label{eq:onshellScalar}
\end{equation}
where the mass $m$ satisfy $\Delta(\Delta-d)=m^2 \ell_{d+1}^2$. To proceed, we break down the computation of boundary correlators in the bulk in two parts. In the near-horizon region, the scalar field couples to the Schwarzian mode, and the corresponding boundary correlators can be computed taking into account their mutual interaction. This is described in Section \ref{subsec:correlatorsNHR}. In the far away region, the spacetime is integrated out as suggested by the holographic renormalization procedure, which is implemented in Section \ref{subsec:matchingUV}. Finally, Section \ref{subsub:anal_structure} is devoted to the analysis of the analytic structure of the retarded Green's function in the complex frequency domain. 
\subsection{Near-horizon region}
\label{subsec:correlatorsNHR}
A dimensional reduction of the scalar field action (\ref{eq:onshellScalar}) using the ansatz (\ref{eq:ansatz}) gives
\begin{equation}
\displaystyle I_{\text{scalar}}=\frac{L^{d-1}}{2}\sum_{\vec{k}}\int_{\mathcal{M}_2}d^2x\sqrt{g^{(2)}}\Phi^{d-1}\left(g^{\mu \nu}\partial_{\mu}\varphi_{\vec{k}}^{*}\:\partial_{\nu}\varphi_{\vec{k}}+\left(\frac{\Phi_0}{\Phi}\right)^{d-2}\left(m^2+\frac{\vec{k}^2}{\Phi^2}\right)|\varphi_{\vec{k}}|^2\right),
\label{eq:action_scalar_2d}
\end{equation}
where the scalar field has been written in a plane wave basis with momentum $\vec{k}$ conjugate to the transverse space coordinates $\vec{y}$, with a scalar amplitude $\varphi_{\vec{k}}(x^{\mu})$ that depends on the two-dimensional coordinates $x^{\mu}$,
\begin{equation}
\varphi(x^{\mu},y^i)=\sum_{\vec{k}}\varphi_{\vec{k}}(x^{\mu})e^{i\vec{k}\,\vec{y}}.
\end{equation}
In the near-horizon region, the dilaton slightly deviates from its extremal value $\Phi=\Phi_0(1+\phi)$,  with $\Phi_0=r_0$ and $\phi \ll 1$ (\ref{eq:linaerized_dilaton}). The action (\ref{eq:action_scalar_2d}) at linear order in $\phi$ is then
\begin{equation}
\begin{split}
\displaystyle I_{\text{scalar}}^{NHR}=&\frac{(r_0 L)^{d-1}}{2}\sum_{\vec{k}}\int_{\mathcal{M}_2}d^2x\sqrt{g^{(2)}}\left(g^{\mu \nu}\partial_{\mu}\varphi_{\vec{k}}^{*}\:\partial_{\nu}\varphi_{\vec{k}}+\left(m^2+\frac{\vec{k}^2}{r_0^2}\right)|\varphi_{\vec{k}}|^2\right)\\
&\:\:\:\:\:\:\:\:\:\:\:+\frac{(r_0 L)^{d-1}}{2}\sum_{\vec{k}}\int_{\mathcal{M}_2}d^2x\sqrt{g^{(2)}}\left((d-1)g^{\mu \nu}\phi\:\partial_{\mu}\varphi_{\vec{k}}^{*}\:\partial_{\nu}\varphi_{\vec{k}}+\left(m^2-\frac{\vec{k}^2}{r_0^2}\right)\phi|\varphi_{\vec{k}}|^2\right)\:.
\end{split}
\label{eq:scalar_interactions}
\end{equation}
As advocated in \cite{Ghosh:2019rcj}, in the near-horizon region, the dilaton is approximately constant $\Phi \approx \Phi_0$ and large for big black holes, so the interaction terms in the second line of (\ref{eq:scalar_interactions}) can be neglected. Therefore, the scalar field $\varphi_{\vec{k}}$ effectively behaves as a free massive field which is decoupled from the linearized dilaton, with an effective constant mass (squared) $m_{eff}^2=m^2+\frac{\vec{k}^2}{r_0^2}$. Given this, we can analyze the boundary correlators of $\varphi_{\vec{k}}$ in the near-horizon region under those circumstances, with the two-dimensional spacetime being described by JT gravity. Fortunately, that task has been addressed in the literature in the past years \cite{Maldacena:2016upp,Bagrets:2017pwq,Mertens:2017mtv,Lam:2018pvp,Bagrets:2016cdf} (see \cite{Mertens:2022irh} for a review).
In fact, the exact euclidean 2-point function of boundary operators in JT gravity with scaling dimension $\Delta_{JT}$ has already been computed in \cite{Mertens:2017mtv,Lam:2018pvp,Bagrets:2016cdf}. It is exact because it accounts for all orders in the Schwarzian mode expansion around the thermal saddle \cite{Mertens:2020pfe}. It was also proven in \cite{Lam:2018pvp,Griguolo:2021zsn,Mertens:2020pfe} that the exact result recovers the 1-loop calculation obtained by \cite{Maldacena:2016upp}.
In the present case, the scaling dimension $\Delta_{JT}$ can be written in terms of $m$ and $\vec{k}$ because of the relation $\Delta_{JT}(\Delta_{JT}-1)=m_{eff}^2\ell_2^2$, yielding$\:$
\begin{equation}
    \Delta_{JT}=\frac{1}{2}+ \sqrt{\frac{1}{4}+(m\ell_2)^2+\frac{\vec{k}^2 \ell_2^2}{r_0^2}}\:.
    \label{eq:dim_NHR}
\end{equation}
The same result (\ref{eq:dim_NHR}) can be obtained by solving the wave equation in the extremal background using the Frobenius method in powers of $r-r_0$ at zero frequency.

The exact euclidean 2-point function can be analytically continued from euclidean boundary time $u$ to real Lorentzian time $t$ via $\tau(u)=u \to -it$ \cite{Mertens:2017mtv,Lam:2018pvp,Mertens:2022irh}. The resultant expression is
\begin{equation}
\langle \mathcal{O}\mathcal{O}\rangle_{\pm}=\frac{1}{Z(\beta)}\prod_{i=1}^2\int_0^{\infty} ds_i^2\:\sinh(2\pi s_i)e^{\pm \frac{i}{2C}(s_1^2-s_2^2)t}e^{-\frac{\beta}{2C}s_2^2}\frac{\Gamma\big(\Delta_{JT} \pm i(s_1 \pm s_2)\big)}{(2C)^{2\Delta_{JT}}\Gamma(2\Delta_{JT})}e^{-s_1^2 \delta}e^{-s_2^2 \delta}\:,
\label{eq:exact_Wightman}
\end{equation}
where $\pm$ stands for the possible orderings appearing in the real-time Wightman functions
\begin{equation}
\langle \mathcal{O}\mathcal{O}\rangle_{+}=\langle \mathcal{O}(t)\mathcal{O}(0)\rangle,\:\:\:\:\:\:\:\:\:\:\:\:\:\langle \mathcal{O}\mathcal{O}\rangle_{-}=\langle \mathcal{O}(0)\mathcal{O}(t)\rangle,
\end{equation}
with a normalization given by $Z(\beta)=\left(\frac{2\pi^2 C}{\beta}\right)^{3/2}e^{\frac{2\pi^2}{\beta}C}$, a scaling dimension $\Delta_{JT}$ (\ref{eq:dim_NHR}), a constant $C=(2\pi M_{\text{gap}})^{-1}$ related to the "mass gap" (\ref{eq:scaleC}) and a regulator $\delta \to 0^+$. The short hand notation indicates $\Gamma(a\pm b \pm c)=\Gamma(a+b+c)\Gamma(a+b-c)\Gamma(a-b+c)\Gamma(a-b-c)$.

\begin{figure}[t]
    \centering
    \includegraphics[width=0.6\textwidth]{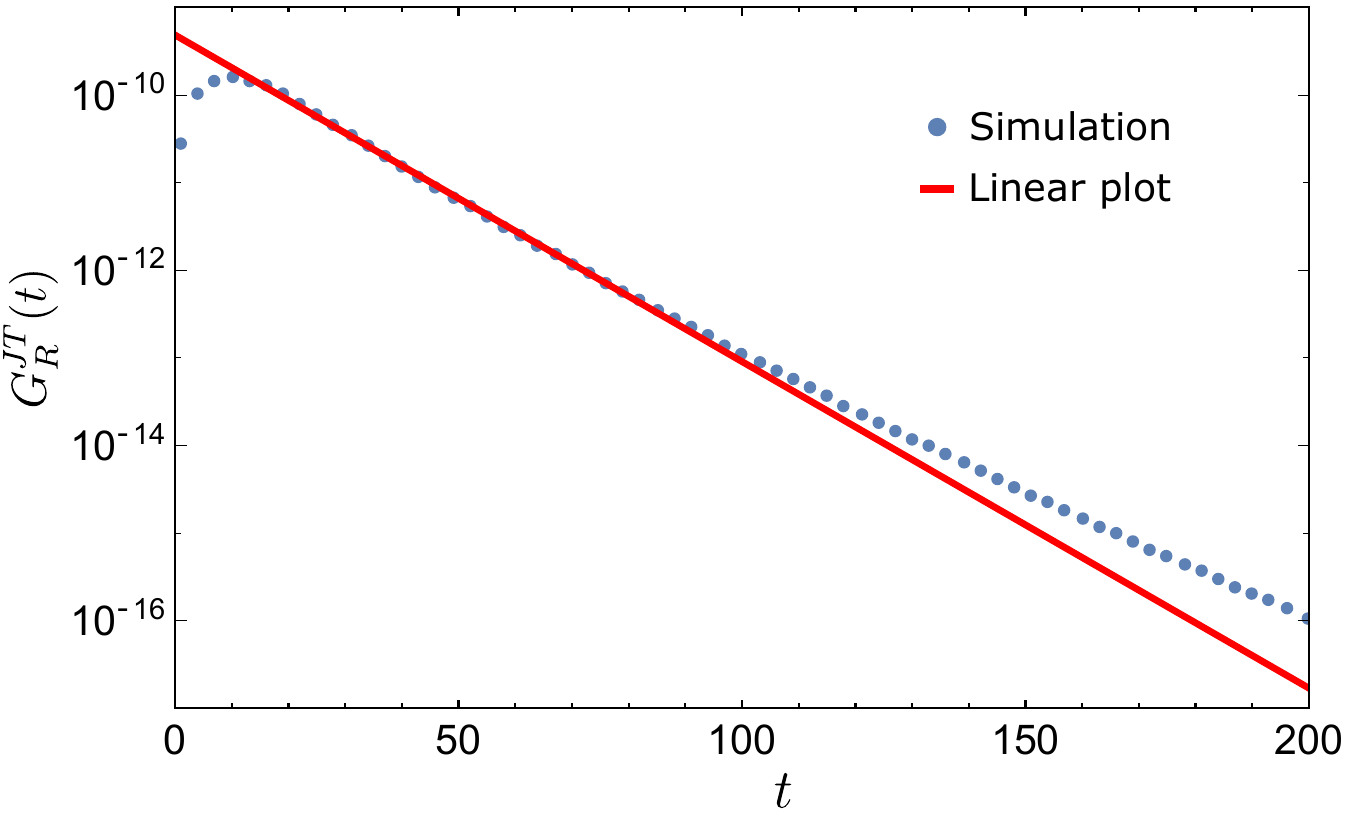}
    \caption{Near-horizon region retarded Green's function $G_R^{JT}(t)$ (\ref{eq:retardedJT}) (in logarithmic scale) vs Lorentzian time $t$. The linear plot in red shows that $G_R^{JT}(t)$ decays exponentially for late Lorentzian times with $t \ll C$. For larger values of $t$ the integrand is highly oscillatory and the numerical integration does not converge. The parameters of the simulation are $\Delta_{JT}=5/4$, $C=100$, $\beta=50$ and $\delta=1/10$.}
    \label{fig:GR_vs_t}
\end{figure}

In the light of the two real-time orderings (\ref{eq:exact_Wightman}), we can also construct the retarded Green's function, which controls the causal response in the boundary theory under a small perturbation \cite{Birmingham:2001pj}. The retarded Green's function is defined as
\begin{equation}
G_R(t)=-i\Theta(t)\langle [\mathcal{O}(t),\mathcal{O}(0)]\rangle=-i\Theta(t)\big(\langle \mathcal{O}\mathcal{O}\rangle_{+}-\langle \mathcal{O}\mathcal{O}\rangle_{-}\big)\:,
\label{eq:defGR}
\end{equation}
where $\Theta(t)$ is the Heaviside step function. Hence, for near-horizon boundary correlators, the exact retarded Green's function is
\begin{equation}
    G^{JT}_R(t)=\frac{2}{Z(\beta)}\prod_{i=1}^2\int_0^{\infty} ds_i^2\:\sinh(2\pi s_i)e^{-\frac{\beta}{2C}s_2^2}\frac{\Gamma\big(\Delta_{JT} \pm i(s_1 \pm s_2)\big)}{(2C)^{2\Delta_{JT}}\Gamma(2\Delta_{JT})}\sin\left(\frac{(s_1^2-s_2^2)}{2C}t\right)e^{-s_1^2 \delta}e^{-s_2^2 \delta}\:.
    \label{eq:retardedJT}
\end{equation}
In Fig. \ref{fig:GR_vs_t} we show a plot of (\ref{eq:retardedJT}) as a function of Lorentzian time $t$. The late time behavior and the analytic structure of (\ref{eq:retardedJT}) are studied in detail below in Section \ref{subsub:anal_structure}. 
\subsection{Matching procedure}
\label{subsec:matchingUV}
\subsubsection{Schwinger-Keldysh formalism}
\label{subsec:SKfieldtheory}
\begin{figure}[t]
    \centering
    \includegraphics[width=0.47\textwidth]{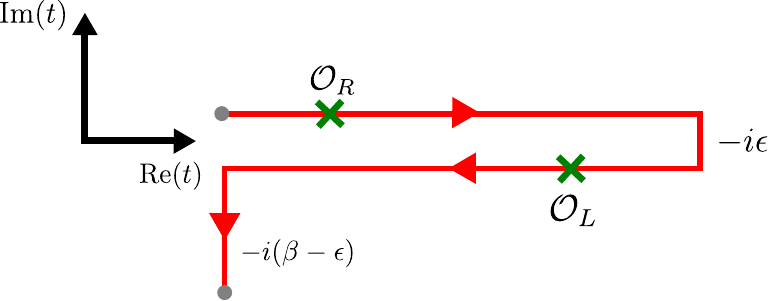}
    \caption{Complex time contour $\mathcal{C}_{SK}$ in the Schwinger-Keldysh formalism for a thermal system.  The Lorentzian sections consist of forward/backward real-time evolution, where the operators $\mathcal{O}_{L/R}$ with sources $J_{L/R}$ are inserted. The Euclidean section has periodicity $\beta=T^{-1}$, which is the inverse of the temperature $T$. The two endpoints of the red contour are identified.}
    \label{fig:SK_contour}
\end{figure}

In this section we briefly review the Schwinger-Keldysh formalism for non-equilibrium field theory (refer to \cite{Liu:2018kfw,Haehl:2016pec} for a broader discussion). Given a system prepared in an initial state, consider the time evolution on a close time path or Schwinger-Keldysh contour $\mathcal{C}_{SK}$ (in Fig. \ref{fig:SK_contour} we show $\mathcal{C}_{SK}$ for an initial thermal state). Real-time observables can then be computed by inserting operators on the Lorentzian parts of the contour. Hence, a central object of this formalism is the generating functional of real-time correlation functions $Z_{SK}[J_L,J_R]$, which depends on the sources $J_{L/R}$ associated to operator insertions $\mathcal{O}_{L/R}$.

A path integral expression of the generating functional $Z_{SK}$ can be given. It can be written in terms of the following ground state $|\Omega\rangle$ expectation value
\begin{equation}
Z_{SK}[J_R,J_L]=\langle \Omega |\mathcal{T}_{\mathcal{C}_{SK}}\:\text{exp}\left(i\int_{\mathcal{C}_{SK}}\Big(\mathcal{L}[\mathcal{O}(x)]+J(x)\mathcal{O}(x)\Big)\right)|\Omega\rangle\:,
\label{eq:SK_generating}
\end{equation}
with Lagrangian $\mathcal{L}$, classical sources $J(x)$ associated to operator insertions $\mathcal{O}(x)$, and integration along the contour $\mathcal{C}_{SK}$ with time-ordering operator $\mathcal{T}_{\mathcal{C}_{SK}}$. 
In addition, it is also convenient to introduce the average-difference basis, which resembles to the implementation of light-cone coordinates, where the operators are
\begin{equation}
\begin{split}
\mathcal{O}_a=\frac{1}{2}(\mathcal{O}_R+\mathcal{O}_L)\:,\:\:\:\mathcal{O}_d=\mathcal{O}_R-\mathcal{O}_L\:,
\end{split}    
\end{equation}
and the source are
\begin{equation}
    J_a=\frac{1}{2}(J_R+J_L)\:,\:\:\:J_d=J_R-J_L\:.
    \label{eq:aver_diff_basis}
\end{equation}
At the level of two-point functions, the real-time retarded Green's function can be obtained by taking functional derivatives with respect to the sources and subsequently setting them to zero
\begin{equation}
    G_R(t_1,t_2)=\frac{(-i)}{Z_{SK}}\frac{\delta^2 Z_{SK}[J_a,J_d]}{\delta J_a(t_1) \delta J_d(t_2)}\Big |_{J=0}\:,
    \label{eq:RetardedSK}
\end{equation}
and similarly for the Keldysh (or symmetric) Green's function
\begin{equation}
G_K(t_1,t_2)=\frac{(-i)}{Z_{SK}}\frac{\delta^2 Z_{SK}[J_a,J_d]}{\delta J_d(t_1) \delta J_d(t_2)}\Big |_{J=0}\:.
\label{eq:KeldyshGreens}
\end{equation}
The expressions (\ref{eq:RetardedSK}) and (\ref{eq:KeldyshGreens}) are not all independent since they are related via the fluctuation/dissipation theorem.
\subsubsection{Planar charged black holes in AdS$_{d+1}$}
\label{subsub:PlanarRN}
\begin{figure}[t]
    \centering
    \includegraphics[width=0.67\textwidth]{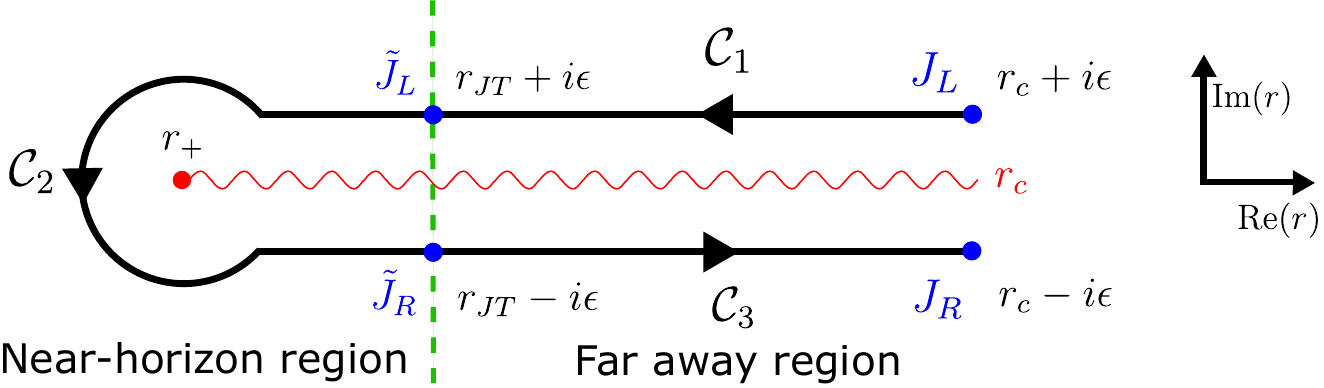}
    \caption{Gravitational Schwinger-Keldysh contour $\mathcal{C}$ in the complex $r$-plane for a constant $v$ slice. The contour winds around the outer horizon with locus at $r=r_+$, encircling the branch cut that runs towards the conformal boundary at $r_c \to \infty$ (in red). It consist of three disjoint parts: $\mathcal{C}=\mathcal{C}_1 \cup \mathcal{C}_2 \cup \mathcal{C}_3$. The contours $\mathcal{C}_1$ and $\mathcal{C}_3$ are part of the far away region and run in opposite directions, at different sides of the branch cut. The contour $\mathcal{C}_2$ is part of the near-horizon region encircling the horizon. The physical sources $J_{L/R}$ are placed at the conformal boundary at $r_c\pm i \epsilon$, whereas the transient sources $\tilde{J}_{L/R}$ are placed at the interface between the near-horizon region and the far away region at $r_{JT}\pm i\epsilon$.}
    \label{fig:grSK_contour}
\end{figure}

The computation of holographic real-time Green's functions can be accomplished by employing the gravitational saddle dual to the Schwinger-Keldysh contour. This labor is given by the dubbed gravitational Schwinger-Keldysh geometry (grSK), motivated from the planar Reissner-Nordström-AdS$_{d+1}$ black hole (\ref{eq:LineEl_Vect}) in ingoing Eddington-Filkenstein coordinates, which is a complex 2-sheeted geometry with \cite{Jana:2020vyx,He:2021jna,He:2022jnc,He:2022deg}
\begin{equation}
    ds^2=-f(r)dv^2+i \beta f(r)dv d\zeta+r^2\delta_{ij}dx^i dx^j\:,\:\:\:\:\:\:\:\:\:\:\mathbf{A}=A_{v}\:dv\:,
    \label{eq:grSK_planarRN}
\end{equation}
where $v$ is the ingoing Eddington–Finkelstein coordinate, $\mathbf{A}$ is the Maxwell 1-form (\ref{eq:LineEl_Vect}), $f(r)$ is the redshift factor (\ref{eq:redshift_vectorP}) and $\beta=T^{-1}$ is the inverse of the Hawking temperature (\ref{eq:temp_chem_pot}). Additionally, $\zeta$ is the complex tortoise coordinate with $dr=\frac{i \beta}{2}fd\zeta$ that follows a contour $\mathcal{C}$ in the complex $r$-plane described in Fig. \ref{fig:grSK_contour}.

A probe scalar field $\varphi(r,v,x^i)$ is introduced into the background geometry (\ref{eq:grSK_planarRN}), which is dual to the double set of boundary operators $\mathcal{O}_{L/R}(t,x^i)$, with  asymptotic behavior related to the physical sources $J_{L/R}$  (that is precisely specified below on (\ref{eq:bdy_condtions_planar_1}) and (\ref{eq:bdy_condtions_planar_3})). The Gaussian scalar action (\ref{eq:onshellScalar}) is modified by replacing the $r$-integral with the complex $\zeta$-integral along the contour $\mathcal{C}$,
\begin{equation}
    I_{\text{scalar}}=\frac{1}{2}\oint_{\mathcal{C}} d\zeta\int dvdx^i\sqrt{-g}(g^{AB}\partial_{A}\varphi\partial_{B}\varphi+m^2\varphi^2)\:.
    \label{eq:onshellScalarSK}
\end{equation}
The holographic AdS/CFT correspondence then asserts that the Schwinger-Keldysh generating functional of correlation functions with sources $J_{L/R}$ (\ref{eq:SK_generating}) is related to the on-shell scalar action (\ref{eq:onshellScalarSK}) via
\begin{equation}
    Z_{grSK}[J_L,J_R]=e^{iI_{\text{scalar}}[J_L,J_R]}\equiv Z_{SK}[J_L,J_R]\:.
    \label{eq:ZgrSK}
\end{equation}
To evaluate (\ref{eq:ZgrSK}) using the viewpoint of the holographic renormalization procedure \cite{Heemskerk:2010hk}, we split the geometry in two parts: the near-horizon region and the far away region. In the near-horizon region, the effective quadratic action in the sources along the contour $\mathcal{C}_2$ is
\begin{equation}
I^{\text{NHR}}_{\text{scalar}}=\int \frac{d\omega}{2\pi}\frac{d^{d-1}k}{(2\pi)^{d-1}}\:\Big(\tilde{J}_a(\omega,\vec{k})\mathcal{I}_{ad}^{JT}(\omega,\vec{k})\tilde{J}_d(-\omega,-\vec{k})+\tilde{J}_d(\omega,\vec{k})\mathcal{I}_{dd}^{JT}(\omega,\vec{k})\tilde{J}_d(-\omega,-\vec{k})\Big),
\end{equation}
where the convention for the Fourier transform is 
\begin{equation}
J(v,\vec{x})=\int \frac{d\omega}{2\pi}\frac{d^{d-1}k}{(2\pi)^{d-1}}J(\omega,\vec{k})e^{-i(\omega v - \vec{k}\vec{x})}.
\end{equation}
The ``transient'' sources $\tilde{J}_{a/d}$ are inserted at the boundary of the near-horizon region, where the labels $a/d$ refer to the average-difference basis (\ref{eq:aver_diff_basis}). The denomination ``transient'' is given because these sources are integrated out at the end of the process, as displayed in (\ref{eq:GeneratingZ_tilde_source}). The two coefficients appearing in the action are the  exact retarded Green's function (\ref{eq:retardedJT}) in momentum space
\begin{equation}
\mathcal{I}^{JT}_{ad}(\omega,\vec{k})=G_{R}^{JT}(\omega,\vec{k})\:,
\end{equation}
together with
\begin{equation}
 \mathcal{I}^{JT}_{dd}(\omega,\vec{k})=\left(n_{\omega}+\frac{1}{2}\right)\mathcal{I}^{JT}_{ad}(\omega,\vec{k})\:,
 \label{eq:I_dd_JT}
\end{equation}
where $n_{\omega}=\frac{1}{e^{\beta \omega}-1}$ is the Boltzmann factor. The relation (\ref{eq:I_dd_JT}) implies that the near-horizon Keldysh Green's function satisfies the fluctuation/dissipation theorem 
\begin{equation}
    G^{JT}_K(\omega,\vec{k})=\frac{1}{2}\left(\mathcal{I}_{dd}^{JT}(\omega,\vec{k})+\mathcal{I}_{dd}^{JT}(-\omega,-\vec{k})\right)=\frac{i}{2}\coth\left(\frac{\beta \omega}{2}\right)\text{Im}\left[G^{JT}_{R}(\omega,\vec{k})\right]\:.
\end{equation} 
Here, we also need to use the property that $G_R(t)$ (\ref{eq:retardedJT}) is real, so that $\big(G_R^{JT}(\omega,\vec{k})\big)^*=G_R^{JT}(-\omega,-\vec{k})$.

In the far away region, we evaluate the action on-shell. To do so, we first solve the scalar wave equation
\begin{equation}
\frac{1}{\sqrt{-g}}\partial_A\left(\sqrt{-g}g^{AB}\partial_B \varphi\right)-m^2\varphi=0\:,
\label{eq:wave_eqn_RN}
\end{equation} 
along with double-Dirichlet boundary conditions: with physical sources $J_{L/R}$ at the asymptotic AdS$_{d+1}$ boundary and with transient sources $\tilde{J}_{L/R}$ at the boundary of the near-horizon region,
\begin{equation}
\text{contour }\mathcal{C}_1:\:\:\:\:\:\varphi_{\omega, {\vec{k}}}\big|_{r_c+ i\epsilon}=J_{L}(\omega,\vec{k}),\:\:\:\:\:\:\:\varphi_{\omega, {\vec{k}}}\big|_{r_{JT}+ i\epsilon}=\tilde{J}_{L}(\omega,\vec{k})\:,
\label{eq:bdy_condtions_planar_1}
\end{equation}
\begin{equation}
\text{contour }\mathcal{C}_3:\:\:\:\:\:\varphi_{\omega, {\vec{k}}}\big|_{r_c- i\epsilon}=J_{R}(\omega,\vec{k}),\:\:\:\:\:\:\:\varphi_{\omega, {\vec{k}}}\big|_{r_{JT}- i\epsilon}=\tilde{J}_{R}(\omega,\vec{k})\:.
\label{eq:bdy_condtions_planar_3}
\end{equation}
In the former expressions $\varphi_{\omega, {\vec{k}}}(r)$ indicates the Fourier amplitude of the scalar field $\varphi(r,v,\vec{x})$. The on-shell action in the far away region along the contour $\mathcal{C}_1 \cup \mathcal{C}_3$ gives two boundary terms with a relative minus sign
\begin{equation}
\begin{split}
    I^{\text{FAR}}_{\text{scalar}}&=\frac{1}{2}\int\frac{d\omega}{2\pi}\frac{d^{d-1}k}{(2\pi)^{d-1}}\left[r^{d-1}f(r)\varphi_{\omega ,{\vec{k}}}(r)\partial_r \varphi^{*}_{\omega , {\vec{k}}}(r)\right]\Big|^{r_c-i\epsilon}_{r_{JT}-i\epsilon}\\
    &-\frac{1}{2}\int\frac{d\omega}{2\pi}\frac{d^{d-1}k}{(2\pi)^{d-1}}\left[r^{d-1}f(r)\varphi_{\omega, {\vec{k}}}(r)\partial_r \varphi^{*}_{\omega, {\vec{k}}}(r)\right]\Big|^{r_c+i\epsilon}_{r_{JT}+i\epsilon},
\end{split}
\label{eq:onshell_FAR_RN}
\end{equation}
where $\varphi^{*}_{\omega , {\vec{k}}} \equiv \varphi_{-\omega ,{-\vec{k}}}$. Finally, the generating functional is obtained after integrating out the transient sources $\tilde{J}_{L/R}$ inserted at the boundary of the near horizon region
\begin{equation}
  Z_{grSK}[J_L,J_R]=\int \mathcal{D}\tilde{J}_L\mathcal{D}\tilde{J}_R\:e^{i(I^{\text{NHR}}_{\text{scalar}}[\tilde{J}_L,\tilde{J}_R]+I^{\text{FAR}}_{\text{scalar}}[J_L,J_R,\tilde{J}_L,\tilde{J}_R])}\:.
  \label{eq:GeneratingZ_tilde_source}
\end{equation}
For a practical illustration of the prior prescription, lets consider a massless scalar field $m=0$ with a trivial spatial momentum dependence $\vec{k}=0$. First of all, we can perform a gradient expansion in frequency \footnote{In three spacetime dimensions, it is possible to solve the wave equation exactly for arbitrary values of mass and momenta. Refer to Section \ref{subsub:RotatingBTZ} below for further details.}
\begin{equation}
\varphi_{\omega,0}(r)=\varphi_{0}(r)+\varphi_{1}(r)\,\omega+\ldots,\:
\end{equation}
and impose the boundary conditions
\begin{equation}
\text{contour }\mathcal{C}_1:\:\:\:\:\:\varphi_{\omega, {0}}\big|_{r_c+ i\epsilon}=J^{L}_0+J^{L}_1 \omega +\ldots,\:\:\:\:\:\:\:\varphi_{\omega, {0}}\big|_{r_{JT}+ i\epsilon}=\tilde{J}^{L}_0+\tilde{J}^{L}_1 \omega +\ldots,
\label{eq:bdy_gradient_L}
\end{equation}
\begin{equation}
\text{contour }\mathcal{C}_3:\:\:\:\:\:\varphi_{\omega, {0}}\big|_{r_c- i\epsilon}=J^{R}_0+J^{R}_1 \omega +\ldots,\:\:\:\:\:\:\:\varphi_{\omega, {0}}\big|_{r_{JT}- i\epsilon}=\tilde{J}^{R}_0+\tilde{J}^{R}_1 \omega +\ldots\:.
\label{eq:bdy_gradient_R}
\end{equation}
At lowest order in the gradient expansion, the scalar  wave equation (\ref{eq:wave_eqn_RN}) is
\begin{equation}
0=\varphi^{''}_{0}(r)+\left(\frac{d-1}{r}+\frac{f'(r)}{f(r)}\right)\varphi^{'}_{0}(r)\:.
\label{eq:eom_gradient}
\end{equation}
The solutions to (\ref{eq:eom_gradient}) with boundary conditions (\ref{eq:bdy_gradient_L}) and (\ref{eq:bdy_gradient_R}) are
\begin{equation}
\text{contour }\mathcal{C}_1:\:\:\:\:\:\varphi_{0}(r+ i \epsilon)=J_{0}^{L}+\left[\frac{(\tilde{J}_{0}^{L}-J_{0}^{L})}{\int_{r_{JT}}^{r_c}du\:\frac{1}{u^{d-1}f(u)}}\right]\int_{r}^{r_c}du\:\frac{1}{u^{d-1}f(u)}\:,
\end{equation}
\begin{equation}
\text{contour }\mathcal{C}_3:\:\:\:\:\:\varphi_{0}(r-i \epsilon)=J_{0}^{R}+\left[\frac{(\tilde{J}_{0}^{R}-J_{0}^{R})}{\int_{r_{JT}}^{r_c}du\:\frac{1}{u^{d-1}f(u)}}\right]\int_{r}^{r_c}du\:\frac{1}{u^{d-1}f(u)}\:.
\end{equation}
The Gaussian integration of the transient sources (\ref{eq:GeneratingZ_tilde_source}) gives the generating functional
\begin{equation}
Z_{grSK}\propto\text{exp}\Bigg\{i\int\frac{d\omega}{2\pi}\frac{d^{d-1}k}{(2\pi)^{d-1}}\Bigg[J^a_0\left(\frac{\mathcal{I}_{ad}^{JT}(\omega,0)}{1+2\kappa\:\mathcal{I}_{ad}^{JT}(\omega,0)}\right)J^d_0+J_0^d\left(\frac{\mathcal{I}_{dd}^{JT}(\omega,0)}{1+2\kappa\:\mathcal{I}_{ad}^{JT}(\omega,0)}\right)J_0^d\Bigg]\Bigg\}\:,
\end{equation}
for $\kappa=\int_{r_{JT}}^{\infty}du\:\frac{1}{u^{d-1}f(u)}$ and $r_c \to \infty$, where the physical sources were expressed in the average-difference basis (\ref{eq:aver_diff_basis}). The constant $\kappa$ can be absorbed in the normalization of the near-horizon boundary operators associated to $\mathcal{I}^{JT}_{ad}=G_R^{JT}$. The holographic retarded Green's function for $\Delta=2$ at leading order is then
\begin{equation}
    G_R(\omega,0)=\frac{\:G_{R}^{JT}(\omega,0)}{1+2\kappa\:G_{R}^{JT}(\omega,0)}+\mathcal{O}(\omega)\:,
    \label{eq:retarded_UV}
\end{equation}
and the holographic Keldysh Green's function at leading order is
\begin{equation}
\begin{split}
G_K(\omega,0)&=\frac{1}{2}\left(\frac{\mathcal{I}_{dd}^{JT}(\omega,0)}{1+2\kappa\:\mathcal{I}_{ad}^{JT}(-\omega,0)}+\frac{\mathcal{I}_{dd}^{JT}(-\omega,0)}{1+2\kappa\:\mathcal{I}_{ad}^{JT}(-\omega,0)}\right)+\mathcal{O}(\omega)\\
&=\frac{i}{2}\coth\left(\frac{\beta \omega}{2}\right)\text{Im}\left[G_R(\omega,0)\right]\:,
\end{split}
\end{equation}
which also satisfies the fluctuation/dissipation theorem. The retarded Green's function (\ref{eq:retarded_UV}) is consistent with the general holographic result for extremal black holes \cite{Faulkner:2009wj}, and with recent calculations regarding near-extremal black holes \cite{Nayak:2018qej,Ghosh:2019rcj} given that the denominator tends to a constant at leading order (refer to (\ref{eq:lowfrec_realxis}) and Fig.~\ref{fig:retardedGPlot}). The structure of (\ref{eq:retarded_UV}) corresponds to a double trace deformation of the effective IR field theory dual to the near-horizon region \cite{Faulkner:2009wj,Heemskerk:2010hk}. Moreover, this matching procedure implies that the retarded Green's function is a rational function of the near-horizon retarded Green's function in momentum space. As a consequence, the analytic structure of (\ref{eq:retarded_UV}) is tightly  connected with the analytic structure of the near-horizon Green's function. We postpone this final discussion for Section \ref{subsub:anal_structure}. 
\subsubsection{Rotating BTZ black holes}
\label{subsub:RotatingBTZ}
In the following, we proceed in a similar vein as in Section \ref{subsub:PlanarRN} for a geometry given by a purely rotating black hole in asymptotically AdS$_3$ (rotating BTZ black hole) \cite{Banados:1992wn}. The principal advantage in this new setup is that the scalar wave equation has exact solutions and the matching procedure can be done for arbitrary values of mass and momenta. Therefore, after noting the rotating BTZ black hole line element in ingoing Eddington-Filkenstein coordinates \cite{Carlip:1995qv}, the corresponding grSK geometry is \cite{Chakrabarty:2020ohe}
\begin{equation}
ds^2=-f(r)dv^2+i \beta f(r)dv d\zeta+r^2\left(d\tilde{\varphi}-\frac{r_- r_+}{r^2}dv\right)^2,\:\:\:\:\:\:\:f(r)=\frac{(r^2-r^2_+)(r^2-r_-^2)}{r^2}\:,
\label{eq:grSK_BTZ}
\end{equation}
where $v$ is the ingoing Eddington–Finkelstein coordinate, $\tilde{\varphi}$ is the angular coordinate spanning $[0,2\pi)$, $\zeta$ is again the complex mock tortoise coordinate with $dr=\frac{i \beta}{2}fd\zeta$ and  Hawking temperature
\begin{equation}
T=\beta^{-1}=\frac{r_+^2-r_-^2}{2\pi r_+}\:,
\end{equation}
for outer/inner horizon radius $r_{\pm}$, respectively, satisfying $f(r_{\pm})=0$. The near-extremal limit $T \to 0$ of this geometry and the effective JT gravity description has been studied in detail in \cite{Ghosh:2019rcj}.
For our purposes, we consider a probe scalar field $\varphi$ in this gravitational background, whose Gaussian dynamic  is also ruled by the Klein-Gordon wave equation (\ref{eq:wave_eqn_RN}) with the present metric (\ref{eq:grSK_BTZ}). To simply the solutions to the wave equation we introduce a new variable $z(r)$, which interpolates between the horizon $r_+$ at $z=0$ and the conformal boundary $r_c\to \infty$ at $z\to 1$,
\begin{equation}
z=\frac{r^2-r_+^2}{r^2-r_-^2}=1-\frac{(r_+-r_-)(r_++r_-)}{r^2-r_-^2}\:,\:\:\:\:\:\:\:\:\: z\in [0,1)\:.
\label{eq:z_r_variables}
\end{equation}
In the text we will often use interchangeably $z \leftrightarrow r$, since they are related by a one-to-one map.

Since the wave equation (\ref{eq:wave_eqn_RN}) is second-order in derivatives, there are two linearly independent solutions referred as the ingoing $G^{in}_{\omega,l}(z)$ and the outgoing $G^{out}_{\omega,l}(z)$  Green's functions. The conventions for the Fourier transform are
\begin{equation}
G(z,v,\tilde{\varphi})=\frac{1}{2\pi}\sum_{l=-\infty}^{\infty}\int \frac{d\omega}{2\pi}G_{\omega, l}(z)e^{-i\omega v}e^{-i l \tilde{\varphi}}.
\end{equation}
These denominations come from their near horizon behavior $z \to 0$
\begin{equation}
G_{\omega, l}^{\text{in}}(z) \sim \text{constant},\:\:\:\:\:\:\:\: G_{\omega, l}^{\text{out}}(z) \sim z^{ik_+}\sim(r-r_+)^{i k_+}\:,
\end{equation}
hence, the ingoing solution is regular at the horizon, whereas the outgoing one is not. We provide more details about these functions in Appendix \ref{app:BTZ}.

A general solution $S_{\omega ,l}(z)$ to the scalar wave equation (\ref{eq:wave_eqn_RN}) in the grSK geometry (\ref{eq:grSK_BTZ}) consists of a linear combination of the ingoing and outgoing functions. In the far away region, we can explicitly write down 
\begin{equation}
\text{contour }\mathcal{C}_1:\:\:\:\:\:S_{\omega ,l}(r+i\epsilon)=A_{\omega ,l}^L\:G_{\omega ,l}^{\text{in}}(r+i\epsilon)+B_{\omega ,l}^L\:G_{\omega ,l}^{\text{out}}(r+i\epsilon)\:,
\end{equation}
\begin{equation}
\text{contour }\mathcal{C}_3:\:\:\:\:\:S_{\omega ,l}(r-i\epsilon)=A_{\omega ,l}^R\:G_{\omega ,l}^{\text{in}}(r-i\epsilon)+B_{\omega ,l}^R\:G_{\omega ,l}^{\text{out}}(r-i\epsilon)\:.
\end{equation}
for some arbitrary constants $A^{L/R}_{\omega ,l}$ and $B^{L/R}_{\omega ,l}$. To simplify the evaluation of the scalar action in the strict limit $r_c \to \infty$, we first inspect the asymptotic behavior towards the AdS$_3$ boundary
\begin{equation}
\text{contour }\mathcal{C}_1:\:\:\:\:\:S_{\omega ,l}(r_c+i\epsilon)\to \tilde{A}_{\omega ,l}^L\:r_c^{-(2-\Delta)}+\tilde{B}_{\omega ,l}^L\: r_c^{-\Delta}+\ldots\:,
\end{equation}
\begin{equation}
\text{contour }\mathcal{C}_3:\:\:\:\:\:S_{\omega ,l}(r_c-i\epsilon)\to \tilde{A}_{\omega ,l}^R\:r_c^{-(2-\Delta)}+\tilde{B}_{\omega ,l}^R\: r_c^{-\Delta}+\ldots\:,
\end{equation}
where $\Delta(\Delta-2)=m^2 \ell_{3}^2$. The linear relation between the pair of constants $(A_{\omega ,l}^{L/R},B_{\omega ,l}^{L/R})$ and $(\tilde{A}_{\omega ,l}^{L/R},\tilde{B}_{\omega ,l}^{L/R})$ follows straightforward from (\ref{eq:AdS3_falloff}),
\begin{equation}
\begin{pmatrix}
\tilde{A}_{\omega ,l}^{L/R}\\
\tilde{B}_{\omega ,l}^{L/R}
\end{pmatrix}=\begin{pmatrix}
a^+_{\omega, l} &a^-_{\omega, l}\\
b^+_{\omega, l}&b^-_{\omega, l}
\end{pmatrix}\begin{pmatrix}
A_{\omega ,l}^{L/R}\\
B_{\omega ,l}^{L/R}
\end{pmatrix}\:,
\label{eq:change_basis_asymptotics}
\end{equation}
for constants $a^{\pm}_{\omega,l}$, $b^{\pm}_{\omega,l}$ defined in (\ref{eq:constant_ap}), (\ref{eq:constant_bp}) and (\ref{eq:constans_abm}). Henceforth, the double-Dirichlet boundary conditions are again imposed using physical sources $J_{\omega,l}^{L/R}$ at the AdS$_3$ boundary $r_c$, and transient sources $\tilde{J}_{\omega,l}^{L/R}$ at the boundary of the near-horizon region $r_{JT}$
 \begin{equation}
\text{contour }\mathcal{C}_1:\:\:\:\:\:\:\:J_{\omega,l}^L\equiv\tilde{A}^L_{\omega, l}\:\:\:(\text{at } r=r_c+i\epsilon)\:\:\:;\:\:\:\:\:\:\:\:\tilde{J}_{\omega,l}^L\equiv S_{\omega,l}(r_{JT}+i\epsilon)\:,
\end{equation}
\begin{equation}
\text{contour }\mathcal{C}_3:\:\:\:\:\:\:\:J_{\omega,l}^R\equiv\tilde{A}^R_{\omega, l}\:\:\:(\text{at } r=r_c-i\epsilon)\:\:\:;\:\:\:\:\:\:\:\:\tilde{J}_{\omega,l}^R\equiv S_{\omega,l}(r_{JT}-i\epsilon)\:.
\end{equation}
Therefore, the pair of constants $(\tilde{A}_{\omega ,l}^{L/R},\tilde{B}_{\omega ,l}^{L/R})$ can be fully determined in terms of $J_{\omega,l}^{L/R}$ and $\tilde{J}_{\omega,l}^{L/R}$, with $\tilde{A}^{L/R}_{\omega ,l}=J^{L/R}_{\omega ,l}$ and
\begin{equation}
\tilde{B}^{L/R}_{\omega,l}=-\left[\frac{a^+_{\omega,l}b^-_{\omega,l}-a^-_{\omega,l}b^+_{\omega,l}}{a^-_{\omega,l}G^{\text{in}}_{\omega,l}(r_{JT})-a^+_{\omega,l}G^{\text{out}}_{\omega,l}(r_{JT})}\right]\tilde{J}_{\omega, l}^{L/R}+\left[\frac{b^-_{\omega,l} G^{\text{in}}_{\omega,l}(r_{JT})-b^+_{\omega,l}G^{\text{out}}_{\omega,l}(r_{JT})}{a^-_{\omega,l}G^{\text{in}}_{\omega,l}(r_{JT})-a^+_{\omega,l}G^{\text{out}}_{\omega,l}(r_{JT})}\right]J_{\omega,l}^{L/R}\:,
\label{eq:tildeB}
\end{equation}
and similarly for $(A_{\omega ,l}^{L/R},B_{\omega ,l}^{L/R})$, by employing the linear transformation (\ref{eq:change_basis_asymptotics}). 

With all these ingredients, we proceed to evaluate the on-shell scalar action (\ref{eq:onshellScalarSK}) along the contour $\mathcal{C}$. In the far away region, the action takes the same expression as for (\ref{eq:onshell_FAR_RN}), yielding
\begin{equation}
\begin{split}
I^{\text{FAR}}_{\text{scalar}}&=\frac{1}{4\pi}\sum_{l=-\infty}^{\infty}\int\frac{d\omega}{2\pi}\Bigg[\Big(rf(r)(S_{\omega ,l}\partial_r S_{-\omega ,-l})\Big)_{r_c+i\epsilon}-\Big(r f(r)(S_{\omega ,l}\partial_r S_{-\omega ,-l})\Big)_{r_{JT}+i\epsilon}\Bigg]\\
&\:\:\:-\frac{1}{4\pi}\sum_{l=-\infty}^{\infty}\int\frac{d\omega}{2\pi}\Bigg[\Big(rf(r)(S_{\omega ,l}\partial_r S_{-\omega -l})\Big)_{r_c-i\epsilon}-\Big(r f(r)(S_{\omega ,l}\partial_r S_{-\omega ,-l})\Big)_{r_{JT}-i\epsilon}\Bigg]\:.
\end{split}
\label{eq:far_action_BTZ}
\end{equation}
The relative minus sign comes from the opposite radial directions of the integration contours $\mathcal{C}_1$ and $\mathcal{C}_3$. A technical remark is that the on-shell action (\ref{eq:far_action_BTZ}) is divergent for $r_c \to \infty$. Hence, we add a counterterm action that works for $1<\Delta<2$, 
\begin{equation}
 I^{ct}_{\text{scalar}}=\frac{1}{4\pi}\sum_{l=-\infty}^{\infty}\int\frac{d\omega}{2\pi} \sqrt{\gamma}(2-\Delta)\Big(S_{\omega ,l}S_{-\omega ,-l}\Big )_{r_c+i\epsilon}-\frac{1}{4\pi}\sum_{l=-\infty}^{\infty}\int\frac{d\omega}{2\pi} \sqrt{\gamma}(2-\Delta)\Big(S_{\omega ,l}S_{-\omega ,-l}\Big )_{r_c-i\epsilon}\:.
 \label{eq:countert_BTZ}
\end{equation}
At the classical level, the Schwarzian is not the leading irrelevant deformation when $1<\Delta_{JT} <\frac{3}{2} $ \cite{Maldacena:2016upp} so this extra constraint should also be imposed, even though it may differ at the quantum level\footnote{I want to thank Pedro J. Martinez for bringing up this point,  as well as the anonymous referee for their comments.}. Then, in the strict $r_c\to \infty$ limit, the sum of (\ref{eq:far_action_BTZ}) and (\ref{eq:countert_BTZ}) is finite
\begin{equation}
\begin{split}
 I^{\text{FAR}}_{\text{scalar}}+ I^{ct}_{\text{scalar}}=&\frac{1}{4\pi}\sum_{l=-\infty}^{\infty}\int\frac{d\omega}{2\pi}\left[(2-2\Delta)\tilde{A}_{\omega, l}^R \tilde{B}_{-\omega, -l}^R-\Big(rf(r)(S_{\omega ,l}\partial_r S_{-\omega ,-l})\Big)_{r_{JT}-i\epsilon}\right]\\
 &\:-\frac{1}{4\pi}\sum_{l=-\infty}^{\infty}\int\frac{d\omega}{2\pi}\left[(2-2\Delta)\tilde{A}_{\omega, l}^L \tilde{B}_{-\omega, -l}^L-\Big(rf(r)(S_{\omega ,l}\partial_r S_{-\omega ,-l})\Big)_{r_{JT}+i\epsilon}\right]\:.
 \end{split}
\end{equation}
The Gaussian integration over the transient sources $\tilde{J}_{L/R}$ inserted at $r_{JT}$ (\ref{eq:GeneratingZ_tilde_source}) gives the coefficients of the resultant action $\mathcal{I}_{ad}$ and $\mathcal{I}_{dd}$ in the average/difference basis (\ref{eq:aver_diff_basis}) for all values of momenta $(\omega,l)$,
\begin{equation}
\begin{split}
\frac{\mathcal{I}_{ad}(\omega,l)}{(1-\Delta)}&=\frac{[b^+_{\omega ,l}G^{\:' in}_{-\omega,-l}(r)-b^-_{\omega ,l}G^{\:' out}_{-\omega,-l}(r)]r f(r)-2[b^+_{\omega ,l}G^{in}_{-\omega,-l}(r)-b^-_{\omega ,l}G^{ out}_{-\omega,-l}(r)]\mathcal{I}_{ad}^{JT}(\omega,l)}{[a^+_{\omega ,l}G^{\:' in}_{-\omega,-l}(r)-a^-_{\omega ,l}G^{\:' out}_{-\omega,-l}(r)]r f(r)-2[a^+_{\omega ,l}G^{in}_{-\omega,-l}(r)-a^-_{\omega ,l}G^{ out}_{-\omega,-l}(r)]\mathcal{I}_{ad}^{JT}(\omega,l)}\Bigg |_{r_{JT}}\\
&\:\:\:\:\:+\frac{b^-_{\omega ,l}G^{\:' in}_{\omega,l}(r)-b^+_{\omega ,l}G^{\:' out}_{\omega,l}(r)}{a^-_{\omega ,l}G^{\:' in}_{\omega,l}(r)-a^+_{\omega ,l}G^{\:' out}_{\omega,l}(r)}\Bigg |_{r_{JT}},
\end{split}
\label{eq:Iad_no_approx}
\end{equation}
as well as,
\begin{equation}
\begin{split}
\frac{\mathcal{I}_{dd}(\omega,l)}{(1-\Delta)}&=\frac{2(a^+_{\omega ,l}b^{-}_{\omega ,l}-a^-_{\omega ,l}b^{+}_{\omega ,l})}{[a^+_{\omega ,l}G^{\:' in}_{-\omega,-l}(r)-a^-_{\omega,l}G^{\:' out}_{-\omega,-l}(r)]}\times\\
&\:\times \frac{[G^{\:' in}_{-\omega,-l}(r)G^{out}_{-\omega,-l}(r)-G^{in}_{-\omega,-l}(r)G^{\:' out}_{-\omega,-l}(r)]\mathcal{I}^{JT}_{dd}(\omega,l)}{[a^+_{\omega ,l}G^{\:' in}_{-\omega,-l}(r)-a^-_{\omega ,l}G^{\:' out}_{-\omega,-l}(r)]r f(r)-2[a^+_{\omega ,l}G^{in}_{-\omega,-l}(r)-a^-_{\omega ,l}G^{ out}_{-\omega,-l}(r)]\mathcal{I}_{ad}^{JT}(\omega,l)}\Bigg |_{r_{JT}}.
\end{split}
\label{eq:Idd_no_approx}
\end{equation}
The results (\ref{eq:Iad_no_approx}) and (\ref{eq:Idd_no_approx}) need to be evaluated at the boundary of the near-horizon region located at $r_{JT}$, which is constrained by $T\ell_2^2\sim r_+-r_- \ll r_{JT}-r_+ \ll r_{JT}$ \cite{Ghosh:2019rcj}. The asymptotic behavior of the ingoing/outgoing solutions in this location gives (refer to Appendix \ref{app:BTZ} for more details)
\begin{equation}
G_{\omega, l}^{\text{in}}\big |_{\partial NHR} \to a^+_{\omega,l}\: \left(\frac{r_{JT}-r_+}{r_+-r_-}\right)^{-\frac{(2-\Delta)}{2}}+b^+_{\omega,l}\: \left(\frac{r_{JT}-r_+}{r_+-r_-}\right)^{-\frac{\Delta}{2}}+\ldots\:,
\end{equation}
\begin{equation}
G_{\omega, l}^{\text{out}}\big |_{\partial NHR} \to a^-_{\omega,l}\: \left(\frac{r_{JT}-r_+}{r_+-r_-}\right)^{-\frac{(2-\Delta)}{2}}+b^-_{\omega,l}\: \left(\frac{r_{JT}-r_+}{r_+-r_-}\right)^{-\frac{\Delta}{2}}+\ldots\:.
\end{equation}
Further, the redshift factor approximates to $f(r_{JT})\approx \frac{(r_{JT}-r_+)^2}{\ell_2^2}$ near extremality (\ref{eq:red_nearext}). Therefore, from (\ref{eq:Iad_no_approx}) we get the final expression for the holographic retarded Green's function
\begin{equation}
G_R(\omega,l)=\tilde{\mathcal{N}}\left[\frac{(\Delta-2)-2\tilde{\kappa}\:G^{JT}_R(\omega,l)}{1+2\tilde{\kappa}\:G^{JT}_R(\omega,l)}\right]\:,
\label{eq:GR_BTZ}
\end{equation}
with $\tilde{\mathcal{N}}=\frac{2(1-\Delta)}{\Delta}[(r_++r_-)(r_{JT}-r_+)]^{\Delta-1}$ and $\tilde{\kappa}=\frac{2\ell_2^2}{\Delta \:r_{JT}(r_{JT}-r_+)}$. Both constants depend on $r_{JT}$, nevertheless they can be absorbed into the normalization of the boundary operators defining $G_R$ and $G_R^{JT}$. The result (\ref{eq:GR_BTZ}) matches with (\ref{eq:retarded_UV}) near-extremality in the limit $\Delta\to 2$ up to a global normalization factor, since
\begin{equation}
\tilde{\kappa}=\frac{\ell_2^2}{r_{JT}(r_{JT}-r_+)}=\int_{r_{JT}}du\:\frac{\ell_2^2}{r_{JT}(u-r_+)^2}\to \int_{r_{JT}}^{\infty}\frac{du}{uf(u)}=\kappa\:.
\end{equation}
Remarkably, the retarded Green's function (\ref{eq:GR_BTZ}) holds for arbitrary momenta $(\omega,l)$ and does not depend on the coefficients $a^{\pm}_{\omega ,l}$, $b^{\pm}_{\omega ,l}$. On the other hand, the holographic Keldysh Green's function obtained from (\ref{eq:Idd_no_approx}) is
\begin{equation}
\begin{split}
G_K(\omega,l)&=\frac{\tilde{\mathcal{N}}}{2}\left[\frac{-2\tilde{\kappa}(\Delta-1)\:\mathcal{I}_{dd}^{JT}(\omega,l)}{1+2\tilde{\kappa}\:\mathcal{I}_{ad}^{JT}(\omega,l)}\right]+\:(\omega,l)\leftrightarrow (-\omega,-l)\\
&=\frac{i}{2}\text{coth}\left(\frac{\beta}{2}(\omega-l\mu_+)\right)\text{Im}[G_R(\omega,l)]\:,
\end{split}
\end{equation}
with chemical potential $\mu_+=\frac{r_-}{r_+}$ conjugated to the black hole angular momentum. The second line follows from the same set of properties detailed in Section \ref{subsub:PlanarRN} by replacing $\beta \omega \to \beta (\omega-l \mu_+)$, and it is the statement of the fluctuation/dissipation theorem in the presence of a chemical potential\cite{Chakrabarty:2020ohe}.
\subsection{Analytic structure}
\label{subsub:anal_structure}
In this section we analyze the late Lorentzian time behavior and the  analytic structure in the complex frequency domain of the  retarded Green's function. First, we  begin with the examination of the exact near-horizon retarded Green's function (\ref{eq:retardedJT}).
For late Lorentzian times $1\ll t \ll C$, the near-horizon retarded Green's function decays exponentially
\begin{equation}
   G_R^{JT}(t) \longrightarrow \frac{1}{\left(\frac{\beta}{\pi}\sinh\left(\frac{\pi}{\beta}t\right)\right)^{2\Delta_{JT}}} \sim e^{-\frac{2\pi \Delta_{JT}}{\beta}t}\:,
   \label{eq:GR_expdecay}
\end{equation}
whereas for later Lorentzian times $C \ll t$ it decays as a power law
\begin{equation}
    G_R^{JT}(t) \longrightarrow \mathcal{N}\frac{1}{t^4}\:,\:\:\:\:\:\: \mathcal{N}=\frac{3\beta^{\frac{5}{2}}}{\pi^{\frac{3}{2}}}\frac{e^{-\frac{2\pi^2 C}{\beta}}}{(2C)^{2\Delta_{JT}-\frac{3}{2}}}\frac{\Gamma^4(\Delta_{JT})}{\Gamma(2\Delta_{JT})}.
    \label{eq:GR_powerlaw}
\end{equation}
The result (\ref{eq:GR_expdecay}) follows from the semiclassical approximation corresponding to the limit $C\to \infty$ \cite{Lam:2018pvp,Mertens:2022irh,Bagrets:2017pwq}. Indeed, the integral expressions of the exact Wightman functions (\ref{eq:exact_Wightman}) are dominated when then integration variables $s_1 \sim s_2 \gg 1$, so that
\begin{equation}
\langle \mathcal{O}\mathcal{O}\rangle_{\pm}=\frac{1}{\left[\frac{\beta}{\pi}\sinh\left(\frac{\pi}{\beta}t\right)\pm i \epsilon\right]^{2\Delta_{JT}}},
\label{eq:Wightman_expdecay}
\end{equation}
hence, yielding (\ref{eq:GR_expdecay}). For Lorentzian times later than $t\gg C$, the integrand of (\ref{eq:exact_Wightman}) becomes highly oscillatory 
\begin{equation}
\sim e^{\pm i f(s_1,s_2)}\:\:\:\:\:\text{with}\:\:\:\:\:\: f(s_1,s_2)=\frac{(s_1^2-s_2^2)}{2C}t.
\end{equation}
Since the saddle point approximation amounts to demand that
\begin{equation}
\nabla_{s} f=\left(\frac{s_1}{C}t,-\frac{s_2}{C}t\right )\approx (0,0)\:,
\label{eq:nablaf}
\end{equation}
in this case the integrand is dominated by contributions with $s_1 ,s_2 \ll 1$. Thus, under those assumptions the integrals decouple
\begin{equation}
\begin{split}
\langle \mathcal{O}\mathcal{O}\rangle_{\pm}&\approx\frac{1}{Z(\beta)(2C)^{2 \Delta_{JT}}}\frac{\Gamma^4(\Delta_{JT})}{\Gamma(2\Delta_{JT})}\left(\int_{0}^{\infty}ds_1^2\:(2\pi s_1) e^{\pm \frac{it}{2C}s_1^2}e^{-\delta s_1^2}\right)\left(\int_{0}^{\infty}ds_2^2\:(2\pi s_2) e^{- \frac{(\beta\pm it)}{2C}s_2^2}e^{-\delta s_2^2}\right)\\
&=\left[\frac{1}{Z(\beta)}\frac{\pi^3 \:\Gamma^4(\Delta_{JT})}{(2C)^{2 \Delta_{JT}-3}\:\Gamma(2\Delta_{JT})}\right]\frac{1}{\big[(2\delta C)-(\pm it)\big]^{3/2}}\frac{1}{\big[(2\delta C)+(\beta \pm it)\big]^{3/2}}\:,
\end{split}
\end{equation}
and the Wightman functions decay as a power law
\begin{equation}
\langle \mathcal{O}\mathcal{O}\rangle_{\pm} \longrightarrow \mathcal{\tilde{N}}\left(\frac{1}{t^3}\pm\frac{i 3\beta}{2t^4}+\ldots\right)\:,\:\:\:\:\:\: \tilde{\mathcal{N}}=\left(\frac{\beta}{\pi}\right)^{3/2}\frac{e^{-\frac{2\pi^2 C}{\beta}}}{(2C)^{2\Delta_{JT}-\frac{3}{2}}}\frac{\Gamma^4(\Delta_{JT})}{\Gamma(2\Delta_{JT})}\:,
\end{equation}
giving the aforementioned result (\ref{eq:GR_powerlaw}). The power law decay is a non-perturbative effect in $C/\beta$ \cite{Mertens:2020pfe}.
This can already be seen at the 1-loop level \cite{Maldacena:2016upp}, where the late Lorentzian time behavior is still exponentially damped. 

\begin{figure}[ht]
    \centering
    \includegraphics[width=\textwidth]{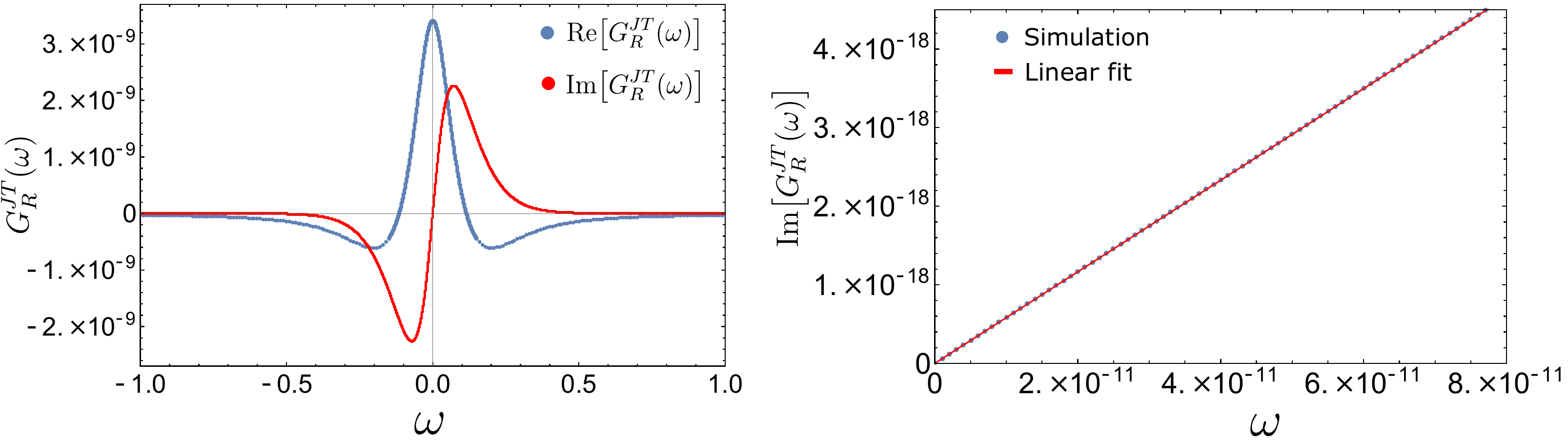}
    \caption{Left: Near-horizon retarded Green's function $G_R^{JT}(\omega)$ as a function of real frequency $\omega$. The real and imaginary parts of $G_R^{JT}(\omega)$ are the blue and red plots, respectively. Right: Imaginary part of the near-horizon retarded Green's function as a function of real frequency $\omega$ for $\omega \ll 1$. The simulated data is presented together with a linear fit. Both simulations were done with parameters $\Delta_{JT}=5/4$, $C=100$, $\beta=50$ and $\delta=1/10$.}
    \label{fig:retardedGPlot}
\end{figure}

The decay (\ref{eq:Wightman_expdecay}) suggests the presence of poles in the lower-half plane of the retarded Green's function in complex  frequency domain, whereas (\ref{eq:GR_powerlaw}) suggests the presence of a branch cut running from the origin. Employing the definition 
\begin{equation}
G_R^{JT}(\omega)=\int_{-\infty}^{\infty} dt\:G_R^{JT}(t)e^{i \omega t}\:,
\end{equation}
the exact close form expression for the upper-half plane $\text{Im}(\omega)>0$ is
\begin{equation}
\begin{split}
G^{JT}_R(\omega)&=\frac{2}{Z(\beta)}\prod_{i=1}^2\int_0^{\infty} ds_i^2\:\sinh(2\pi s_i)e^{-\frac{\beta}{2C}s_2^2}\frac{\Gamma\big(\Delta_{JT} \pm i(s_1 \pm s_2)\big)}{(2C)^{2\Delta_{JT}-1}\Gamma(2\Delta_{JT})}\frac{(s_1^2-s_2^2)}{(s_1^2-s_2^2)^2-4C^2\omega^2}e^{-s_1^2 \delta}e^{-s_2^2 \delta}\\
&=2\prod_{i=1}^2\int_0^{\infty} ds_i\:F(s_1,s_2)\:\frac{a(s_1,s_2)}{a(s_1,s_2)^2-\omega^2}\:.
\end{split}
\label{eq:GR_upper_half}
\end{equation}
For later convenience, we have defined
\begin{equation}
F(s_1,s_2)=\frac{4s_1s_2}{Z(\beta)}\sinh(2\pi s_1)\sinh(2\pi s_2)e^{-\frac{\beta}{2C}s_2^2}\frac{\Gamma\big(\Delta_{JT} \pm i(s_1 \pm s_2)\big)}{(2C)^{2\Delta_{JT}}\Gamma(2\Delta_{JT})}e^{-s_1^2 \delta}e^{-s_2^2 \delta}\:,
\end{equation}
and
\begin{equation}
a(s_1,s_2)=\frac{(s_1^2-s_2^2)}{2C}\:.
\end{equation}
The exact Green's function in the upper-half plane (\ref{eq:GR_upper_half}) generalizes the 1-loop result studied in detail in \cite{Qi:2018rqm}.
Besides, (\ref{eq:GR_upper_half}) can be further analytically continued to the $\omega$ real line. Since $G_R^{JT}(t)$ (\ref{eq:retardedJT}) is real, the real/imaginary parts of $G_R^{JT}(\omega)$ are even/odd functions of $\omega$ real, respectively. Thus, for simplicity we proceed to perform the analytic continuation towards $\omega \geq 0$. We can use the identity
\begin{equation}
\displaystyle\frac{a}{a^2-\omega^2}=\frac{C}{2\sigma_+(s_2)}\left(\frac{1}{s_1-\sigma_+(s_2)}-\frac{1}{s_1+\sigma_+(s_2)}\right)+\frac{C}{2\sigma_-(s_2)}\left(\frac{1}{s_1-\sigma_-(s_2)}-\frac{1}{s_1+\sigma_-(s_2)}\right),
\end{equation}
to isolate the single poles in $s_1$ that parametrically depend on $s_2$ and $\omega$, where
\begin{equation}
\sigma_{\pm}(s_2)=\sqrt{s_2^2\pm2C\omega}.
\end{equation}
To approach the real frequency line from the upper-half plane we use the following identity for every single pole
\begin{equation}
\frac{1}{x\pm i \eta}=\text{P.V.}\left(\frac{1}{x}\right)\mp i \pi \delta(x),\:\:\:\:\:\:\:\:\:\:\:\: \eta \to 0^{+}\:,
\end{equation}
where P.V. stands for "principal value" and $\delta(x)$ is the Dirac delta function. Therefore, for real $\omega \geq 0$, the exact near-horizon retarded Green's function is
\begin{equation}
\begin{split}
G_R^{JT}(\omega)&=2\int_{0}^{\infty}ds_2\:\left[\text{P.V.}\left(\int_{0}^{\infty}ds_1\:F(s_1,s_2)\:\frac{a(s_1,s_2)}{a(s_1,s_2)^2-\omega^2}\right)\right]\\
&\:\:\:\:\:\:+i \pi C\left[\int_{0}^{\infty}ds_2\:\frac{F(\sigma_{+}(s_2),s_2)}{\sigma_{+}(s_2)}-\int_{\sqrt{2C \omega}}^{\infty}ds_2\:\frac{F(\sigma_{-}(s_2),s_2)}{\sigma_{-}(s_2)}\right]\:.
\end{split}
\label{eq:GR_real_line}
\end{equation}
A notable property of (\ref{eq:GR_real_line}) is that its imaginary part, which is given by the second line of (\ref{eq:GR_real_line}), is written in terms of a single integral. With such a simplification, it is possible to plot $G_R^{JT}(\omega)$ for $\omega$ real, as portrayed in Fig. \ref{fig:retardedGPlot}. We opted to use the Kramers-Kronig relation to compute the real part of the retarded Green's function
\begin{equation}
\text{Re}(G^{JT}_R(\omega))=\text{P.V.}\left(\frac{2}{\pi}\int_0^{\infty} d\omega '\:\left(\frac{\omega '\text{Im}(G_R^{JT}(\omega '))}{\omega '+\omega}\right)\frac{1}{\omega '-\omega}\right)\:.
\end{equation} 
Furthermore, the close form low frequency expression of the imaginary part turns out to scale linearly with frequency (see right plot of Fig. \ref{fig:retardedGPlot}),
\begin{equation}
\begin{split}
\text{Im}(G_R^{JT}(\omega))&=\omega\Bigg(\frac{ \Gamma^2(\Delta)}{\Gamma(2\Delta)}\frac{2\pi}{Z(\beta)(2C)^{2(\Delta-1)}}\int_{0}^{\infty}ds_2\:|\Gamma(\Delta+i2s_2)|^2\sinh(2\pi s_2)e^{-(\frac{\beta}{2C}+2\delta)s_2^2}\times\\
&\:\:\:\:\:\:\:\:\:\:\:\times \Big[2\pi\cosh(2\pi s_2)+\sinh(2\pi s_2)\big(i\psi(\Delta+2is_2)-i\psi(\Delta-2is_2)-2s_2\delta\big)\Big]\Bigg),
\end{split}
\label{eq:lowfrec_realxis}
\end{equation}
where $\psi(x)=\frac{\Gamma'(x)}{\Gamma(x)}$ is the digamma function. 

The close form expression of the exact retarded Green's function in the lower-half plane is still unknown. Because of that, in Section \ref{subsub:ToyModel} we present a toy model for the retarded Green's function that has a close form Fourier transform and also captures the late Lorentzian time behavior (\ref{eq:GR_expdecay}) and (\ref{eq:GR_powerlaw}). There, the existence of a branch cut can be explicitly discerned due to the presence of a non-meromorphic function.

The discussion so far has gravitated around the near-horizon retarded Green's function $G_R^{JT}(\omega)$. As it was discussed in Section \ref{subsec:matchingUV}, the holographic retarded Green's function $G_R$ is a rational function of the near-horizon result $G_R^{JT}$ in momentum space. The direct implication of this matching procedure is that the holographic retarded Green's function will also have poles, and particularly, a branch cut in the lower half-plane for generic values of $\Delta$. For instance, if $G_R^{JT}$ has a simple pole at $\omega=\omega_0$, then  $G_R^{JT}\sim \frac{c}{\omega-\omega_0}$ and
\begin{equation}
G_R\sim\frac{(\Delta-2)(\omega-\omega_0)-2\tilde{k}c}{\omega-\omega_0+2\tilde{k}c}
\end{equation}
will have a pole if $\Delta \neq 1$ at a different location where $\omega=\omega_0-2\tilde{k}c$.

Physically, this means that at low temperatures these novel quantum effects induce a slower thermalization rate, since the  Green's function decay is no longer exponentially damped for $t \gtrsim C$. However, recall that in the planar black hole geometry, the transverse space was compactified on a torus with characteristic length $L$. In the decompactification limit, $L \to \infty$, so that the effective coupling $C\to \infty$ (\ref{eq:scaleC}). The presence of quantum effects appearing at low temperature is thus sensitive to the order of limits $T\to 0$ and $L \to \infty$. Taking $L \to \infty $ first, suppresses  the quantum effects before reaching the $T\to 0$ regime, while taking $T \to 0$ first, keeps manifest the quantum effects. 
\subsubsection{Toy model}
\label{subsub:ToyModel}
A toy model (TM) for the near-horizon retarded Green's function that captures the late Lorentzian time decay (\ref{eq:GR_expdecay}) and (\ref{eq:GR_powerlaw}) is given by
\begin{equation}
    \displaystyle G^{TM}_R(t)=\Theta(t)\left(\frac{1}{\left(\frac{\beta}{\pi}\sinh\left(\frac{\pi}{\beta}t\right)\right)^{2\Delta_{JT}}}+\mathcal{J}(t,C)\frac{1}{t^4}\right),
    \label{eq:GR_TM}
\end{equation}
where $\mathcal{N}$ is the constant (\ref{eq:GR_powerlaw}), $\Theta(t)$ is the Heaviside theta function, and $\mathcal{J}$ is a function that smoothly interpolates between $\mathcal{J}(t\leq 0,C)=0$ and $\mathcal{J}(t\geq C,C)=\mathcal{N}$ with
\begin{equation}
\mathcal{J}(t,C)=\mathcal{N}\left[\Theta(-t+C)S_4(t/C)+\Theta(t-C)\right],
\end{equation}
where $S_4(t)$ is the 9-th order smoothstep function\footnote{The $n$-th order smoothstep function $S_{N}(t)$ , with $n$ odd and $N=\frac{n-1}{2}$, is a Hermite interpolation of degree $n$ between $[0,1]$ for $t\in[0,1]$. The mininum $n$ such that $S_N(t)\frac{1}{t^4}\to 0$ when $t\to 0$ is $n=9$.}\begin{equation}
S_4(t)=70 t^9 - 315 t^8 + 540 t^7 - 420 t^6 + 126 t^5\,.
\end{equation}
The close form Fourier transform of (\ref{eq:GR_TM}) consist of three parts
\begin{equation}
G_R^{TM}(\omega)=G^I(\omega)+G^{II}(\omega)+G^{III}(\omega)\:.
\end{equation}
The first piece is meromorphic and comes from the hyperbolic function \cite{Son:2002sd}
\begin{equation}
G^{I}(\omega)\sim \left|\Gamma\left(\frac{\Delta_{JT}}{2}-\frac{i \beta \omega}{4\pi}\right)\right|^4 \Bigg(\csc(\pi \Delta_{JT})-\cot(\pi \Delta_{JT})\cosh\left(\frac{\beta \omega}{2}\right)+i\sinh\left(\frac{\beta \omega}{2}\right)\Bigg),
\end{equation}
which has poles at $\omega=-2i\pi T (\Delta_{JT}+2n)$ for $n \in \mathbb{N}_0$. The second piece is also meromorphic and comes from the early time power law decay with $t\in [0,C]$,
\begin{equation}
\begin{split}
G^{II}(\omega)&=\mathcal{N}\Big[\frac{1}{\omega^6 C^9}\big(-8400 - i7560 \omega C  + 3240 \omega^2C^2 + 
 i840  \omega^3C^3 - 126  \omega^4C^4\big)\\
 &\:\:\:\: +\frac{e^{i \omega C}}{\omega^6 C^9}\big(8400 - i840   \omega C+ 120 \omega^2C^2 + i20  \omega^3C^3 - 
 4 \omega^4C^4 -  i \omega^5C^5\big)\Big].
 \end{split}
\end{equation}
Finally, the third piece is non-meromorphic and comes from the late time power law decay with $t \in [C,\infty)$,
\begin{equation}
G^{III}(\omega)=\mathcal{N}\left[-\frac{e^{i \omega C}}{6C^3}\big((\omega C)^2-i\omega C-2\big)+\frac{\pi \omega^3}{12}+\frac{i \omega^3}{6}Ci(\omega C)-\frac{ \omega^3}{6}Si(\omega C)\right]\:.
\end{equation}
The functions $Si(z)$ and $Ci(z)$ are the Sine and Cosine integrals \cite{NIST:DLMF}, which are meromorphic and non-meromorphic functions, respectively. The Cosine integral has a branch cut that runs from the origin $z=0$ to infinity. In the semiclassical limit $C \to \infty$ the constant $\mathcal{N}\to 0$ and the branch cut disappear. Moreover, because $\mathcal{N}\sim e^{-\frac{2\pi^2 C}{\beta}}$ the branch cut turns out to be a non-perturbative hallmark of the quantum effects coming from the near-horizon region at low temperatures.
\section{Discussion}
\label{sec:discussion}
We have studied holographic real-time correlators at finite temperature for arbitrary dimensions by working with a probe scalar field in a charged planar near-extremal black hole in asymptotically AdS spacetime. The novel features arising at low temperatures from the quantum fluctuations of the Schwarzian mode are encoded in the exact near-horizon retarded Green's function derived in (\ref{eq:retardedJT}). Using a matching procedure in the grSK geometry, we were able to relate the near-horizon result with the full retarded Green's function in momentum space as concluded in (\ref{eq:retarded_UV}) and (\ref{eq:GR_BTZ}). Moreover, we studied the analytic structure of the retarded Green's function and presented exact expressions for the upper half-plane (\ref{eq:GR_upper_half}) and the real frequency line (\ref{eq:GR_real_line}). We argued, using a toy model (\ref{eq:GR_TM}), for the existence of a branch cut in the lower half-plane, and stated that it is a non-perturbative hallmark that prevails as long as the planar transverse space is kept compact.

Future research ought to generalize the holographic renormalization formalism in the grSK geometry, to account for higher point real-time correlation functions. The study of higher point functions in the grSK geometry was first considered in \cite{Jana:2020vyx}, where the authors dealt with a self-interacting scalar field in the bulk. Additionally, future generalizations of the present work consist of computing correlators of conserved currents that incorporate quantum effects coming from the Schwarzian mode.

To conclude, a complementary outlook of our results comprises of the correlators as a proxy for computational purposes, since the main takeaways might also apply for black holes in asymptotically flat spacetime. Hence, the search for new features in observables associated to astrophysical near-extremal black holes (near-extremal Kerr black holes) provides an appealing direction to pursue in the future.
\section*{Acknowledgments}
I especially thank Mukund Rangamani for suggesting the problem and for fruitful discussions. In addition, I want to thank Pedro J. Martinez for giving feedback on the final draft, and the anonymous referee for their comments. I also thank Victor Godet, Giuseppe Policastro, Balt C. van Rees and Julio Virrueta for discussions, and Sophia Minnillo for proofreading the draft. LD was supported by U.S. Department of Energy grant DE-SC0020360 under the HEP-QIS QuantISED program, and by a Dean's Distinguished Graduate Fellowship from the College of Letters and Science of the University of California, Davis. LD appreciates the hospitality at the Center for Quantum Mathematics and Physics (QMAP), Davis, as well as at the International Center for Interdisciplinary Science and Education (ICISE) at Quy Nhon, Vietnam in the context of the ``Advanced Summer School in Quantum Field Theory and Quantum Gravity''.

\appendix

\section{Supplement: black hole geometry}
\label{app:planarRNapp}
In this Appendix we describe additional material that complements Section \ref{sec:Planar_RN}. Included here are Section \ref{subsec:geom_low_T}, which is devoted to a detailed description of the extremal geometry and the near-extremal geometry. In addition, Section \ref{subsec:dim_reduction} deals with the dimensional reduction procedure for the near-extremal black hole geometry. Other references describing the Reissner-Nordström geometry and JT gravity include \cite{Iliesiu:2020qvm,Nayak:2018qej,Moitra:2019bub,Sachdev:2019bjn}.
\subsection{Geometry at low temperatures}
\label{subsec:geom_low_T}
\subsubsection{Extremal geometry}
\label{subsubsec:ext_geom}
The extremal geometry is defined to have zero temperature $T=0$. In this case, the inner and outer horizons coincide $r_0\equiv r_+=r_-$. Moreover, the parameters $q$ and $m$ are fully determined as functions of $r_0$
\begin{equation}
q^2=\frac{d }{(d-2)}\frac{r_0^{2d-2}}{\ell_{d+1}^2},\:\:\:\:\:\:\:\:\:\:\:m=\frac{2(d-1)}{(d-2)}\frac{r_0^d}{\ell_{d+1}^2}\:,
\label{eq:Ext_mass_charge}
\end{equation}
as well as  the Maxwell vector field and chemical potential
\begin{equation}
A_{\tau}=-i \mu \left(1-\frac{r_0^{d-2}}{r^{d-2}}\right),\:\:\:\:\:\:\:\:\:\: \mu=\sqrt{\frac{d-1}{2(d-2)}}\frac{q}{r_0^{d-2}}\:.
\end{equation}
At extremality, the geometry can be divided in two regions: the near-horizon region and the far away region.

The near-horizon region is given by $r-r_0 \ll r_0$. In this regime, the geometry develops an infinite throat so that the topology becomes AdS$_2\times T^{d-1}$. This can be explicitly seen because the redshift factor is approximated as
\begin{equation}
f(r)=\frac{(r-r_0)^2}{\ell_2^2}+\mathcal{O}\big((r-r_0)^3\big)\:,
\label{eq:red_ext}
\end{equation}
and the following scaling
\begin{equation}
    z=\frac{\lambda \ell_2^2}{r-r_0},\:\:\:\:\:\:\: \tilde{t}=\lambda \tau,\:\:\:\:\:\: \lambda \to 0\:,
\end{equation}
for finite $z,\tilde{t}$ yields
\begin{equation}
    ds^2=\frac{\ell_2^2}{z^2}(d\tilde{t}\,^2+dz^2)+r_0^2\delta_{ij}dx^idx^j\:.
\end{equation}
The induced AdS$_2$ length scale is $\ell_2=\frac{\ell_{d+1}}{\sqrt{d(d-1)}}$, and the asymptotic AdS$_2$ region is located at $\ell_2 \ll r-r_0$. Since we are going to deal with big black holes, i.e. $\ell_{d+1}\ll r_0$, the asymptotic AdS$_2$ region is constrained by $\ell_{d+1}\sim \ell_2 \ll r-r_0 \ll r_0$.

On the other hand, the far away region is given by $r-r_0 \gg r_0$. It overlaps with the near-horizon region at surfaces with $r=r_{JT}$ provided that $\ell_2 \ll r_{JT}-r_0 \ll r_0$ (see Fig. \ref{fig:sketchBH}). The subscript JT will become more transparent when describing the effective gravitational dynamics in Section \ref{subsec:effectiveJT}.
\subsubsection{Near-extremal geometry}
\label{subsubsec:nearext_geom}
Near-extremal black holes have non zero temperature, low enough such that the near-horizon geometry slightly deviates from AdS$_2\times T^{d-1}$. To study the near-extremal regime, we can think of increasing the temperature $T$ of the black (or equivalently, its mass) while keeping the charge $q(r_0)$ fixed as in (\ref{eq:Ext_mass_charge}). Working in the canonical ensemble, parameterized by the temperature and the charge $(T,q(r_0))$, amounts to fixing the field strength on the boundary of AdS$_{d+1}$. The exact condition for a big black hole to be near-extremal is
\begin{equation}
    T\ell_2^2 \ll \ell_2 \ll r_0\:.
    \label{eq:Temp_cond}
\end{equation}
Under this assumption, the redshift factor can be approximated in the near-horizon region $r-r_0 \ll r_0$ as
\begin{equation}
f(r)=\frac{(r-r_-)(r-r_+)}{\ell_2^2}+\ldots=\frac{(r-r_0)^2-(2\pi \ell_2^2 T)^2}{\ell_2^2}+\ldots\approx \frac{(r-r_0)^2}{\ell_2^2}\,.
\label{eq:red_nearext}
\end{equation}
Thus, for low enough temperatures such that (\ref{eq:Temp_cond}) and (\ref{eq:red_nearext}) hold, there are hypersurfaces at $r=r_{JT}$ such that
\begin{equation}
T\ell_2^2 \ll \ell_2 \ll r_{JT}-r_0\ll r_0\:,
\label{eq:location_rJT}
\end{equation}
where the near-horizon region and the far away region overlap. In the latter region, the geometry behaves locally as vacuum AdS$_2$ since (\ref{eq:red_nearext}) approximates to (\ref{eq:red_ext}). In this way, the boundary Schwarzian mode induced by the effective JT gravity description discussed in Section \ref{subsec:effectiveJT} is located at $r=r_{JT}$. 
\subsection{Dimensional reduction}
\label{subsec:dim_reduction}
In this Appendix we explain in depth a way to obtain the effective JT gravity action in the s-wave sector (\ref{eq:actiondimRed}) starting from the ansatz (\ref{eq:ansatz}) and (\ref{eq:MaxwellSwave}), which we recast here again
\begin{equation}
    ds^2=g_{AB}dx^Adx^B=g_{\mu \nu}(x^{\rho})dx^{\mu}dx^{\nu}+\Phi^2(x^{\rho}) \delta_{ij}dy^i dy^j\:,
    \label{eq:ansatzApp}
\end{equation}
with
\begin{equation}
A_{M}=(A_{\mu}(x^{\rho}),0)\:.
\end{equation}
An incomplete list of similar calculations in the literature is \cite{Iliesiu:2020qvm,Nayak:2018qej,Moitra:2019bub,Ghosh:2019rcj,Davison:2016ngz,Sachdev:2019bjn,Banerjee:2021vjy,Banerjee:2023quv}. To begin with, recall the following identity relating the Ricci scalar $R^{(d+1)}$ computed using the metric $g_{AB}$ and the Ricci scalar $R^{(2)}$ computed using the metric $g_{\mu \nu}$ 
\begin{equation}
\begin{split}
R^{(d+1)}&=R^{(2)}-2e^{-(d-1)\sigma}\nabla^{\mu}\nabla_{\mu}e^{(d-1)\sigma}+(d-1)(d-2)\nabla_{\mu}\sigma\nabla^{\mu}\sigma,
\end{split}
\label{eq:RicciFull}
\end{equation}
where the dilaton can be rewritten as $\Phi(x)=e^{\sigma(x)}$. Furthermore, the trace of the extrinsic scalar curvature $K^{(d)}$ computed with respect to the $d$-dimensional boundary metric $\gamma^{(d)}_{AB}$, is related to the trace of the extrinsic curvature $K^{(1)}$ computed with respect to the $1$-dimensional boundary metric $\gamma^{(1)}_{\mu \nu}$ via 
\begin{equation}
K^{(d)}=K^{(1)}+\frac{(d-1)}{2\Phi^2}n^{\mu}\partial_{\mu}\Phi^2,
\end{equation}
where $n^{\mu}$ is a unit-normal vector outwardly directed. Before evaluating the on-shell  gravitational action (\ref{eq:EHM_action}), we also need to Weyl rescale the two-dimensional metric as
\begin{equation}
g_{\mu\nu}(x)\to \left(\frac{\Phi_0}{\Phi}\right)^{d-2}g_{\mu \nu}(x),
\end{equation}
for a constant $\Phi_0$. As a consequence, the following transformations yield
\begin{equation}
R^{(2)}\to  \left(\frac{\Phi}{\Phi_0}\right)^{d-2}\left(R^{(2)}+(d-2)\nabla^{\mu}\nabla_{\mu}\sigma\right)\:,
\end{equation}
\begin{equation}
K^{(1)}\to \left(\frac{\Phi}{\Phi_0}\right)^{\frac{d-2}{2}}\left(K^{(1)}+\frac{d}{2\Phi}n^{\mu}\partial_{\mu}\Phi\right)\:.
\end{equation}
We can replace the identities in the gravitational action (\ref{eq:EHM_action}) to obtain
\begin{equation}
\begin{aligned}
I&=-\frac{L^{d-1}}{16\pi G_N^{(d+1)}}\int_{\mathcal{M}_2} d^2x\sqrt{g^{(2)}}\left(\Phi^{d-1}R^{(2)}+\frac{d(d-1)}{\ell_{d+1}^2}\Phi_0^{d-2}\Phi-\frac{\Phi^{2d-3}}{\Phi_0^{d-2}}F_{\mu \nu}F^{\mu \nu}\right)\\
&\:\:\:\:\:\:\:\:-\frac{L^{d-1}}{8\pi G_N^{(d+1)}}\int_{\partial\mathcal{M}_2} d\tau \sqrt{\gamma^{(1)}}\left(\Phi^{d-1}K^{(1)}\right)\:.
\end{aligned}
\end{equation}
The kinetic dilaton term cancels out with the extra boundary contribution arising from the trace of the scalar curvature term. After integrating out the gauge fields, as discussed in \cite{Ghosh:2019rcj,Iliesiu:2020qvm,Moitra:2019bub}, the two-dimensional Einstein-dilaton action is obtained
\begin{equation}
I=-\frac{L^{d-1}}{16\pi G_N^{(d+1)}}\left[\int_{\mathcal{M}_2} d^2x\sqrt{g^{(2)}}\left(\Phi^{d-1}R^{(2)}-U(\Phi)\right)+2\int_{\partial \mathcal{M}_2}d\tau\:\sqrt{\gamma^{(1)}}\:\Phi^{d-1}K^{(1)}\right]\:,
\label{eq:Einstein_dilaton}
\end{equation}
with a dilaton potential
\begin{equation}
U(\Phi)=-\frac{d(d-1)}{\ell_{d+1}^2}\Phi_0^{d-2}\Phi+(d-1)(d-2)q^2 \frac{\Phi_0^{d-2}}{\Phi^{2d-3}}\:,
\end{equation}
for a sector of fixed charge $q(r_0)$ (\ref{eq:Ext_mass_charge}). The equations of motion inferred from (\ref{eq:Einstein_dilaton}) after varying $\delta \Phi$ and $\delta g^{\mu \nu}$, respectively, are
\begin{equation}
(d-1)\Phi^{d-2}R-U'(\Phi)=0\:,
\label{eq:eom1}
\end{equation}
\begin{equation}
\nabla_{\mu}\nabla_{\nu} (\Phi^{d-1})-g_{\mu \nu}\left(\nabla_{\sigma}\nabla^{\sigma}(\Phi^{d-1})+\frac{1}{2}U(\Phi)\right)=0\:.
\label{eq:eom2}
\end{equation}
A solution to them is given by
\begin{equation}
ds^2_{(2d)}=g_{\mu\nu}dx^{\mu}dx^{\nu}=\left(\frac{\Phi(r)}{\Phi_0}\right)^{d-2}\left(f(r)d\tau^2+\frac{dr^2}{f(r)}\right),\:\:\:\:\:\:\:\: \Phi(r)=r\:,
\label{eq:sol_EinsDil}
\end{equation}
with redshift factor $f(r)$ as in (\ref{eq:redshift_vectorP}) for arbitrary $m$ and $q$.
Note that the parameter $\Phi_0$ is determined from (\ref{eq:sol_EinsDil}) to be $\Phi_0=r_0$. Hence, the solution to the equations of motion (\ref{eq:eom1}) and (\ref{eq:eom2}) corresponds to the s-wave truncation of the Reissner-Nordström-AdS$_{d+1}$ saddle described in (\ref{eq:LineEl_Vect}) and (\ref{eq:redshift_vectorP}). 

With this effective two-dimensional  Einstein-dilaton theory at hand, in the same manner as in Appendix \ref{subsec:geom_low_T}, we can split the geometry in two parts: the near-horizon region and the far away region.
\subsubsection{Near-horizon region}
\label{subsubsec:NHR}
In the near-horizon region $r-r_0 \ll r_0$, in order to account for deviations away from extremality where $\Phi=\Phi_0$, we proceed to linearly expand out the dilaton around $\Phi_0$ such that
\begin{equation}
\Phi=\Phi_0(1+\phi(r))\:\:,\:\:\:\:\:\:\phi \ll 1\:.
\label{eq:linaerized_dilaton}
\end{equation}
Thus, the effective action in the near-horizon region is
\begin{equation}
\begin{split}
I_{NHR}&=-\frac{L^{d-1}\Phi_0^{d-1}}{16\pi G_N^{d+1}}\left[\int_{\mathcal{M}_{NHR}}d^2x\sqrt{g^{(2)}}R^{(2)}+2\int_{\partial \mathcal{M}_{NHR}}dx\sqrt{\gamma^{(1)}}K^{(1)}\right]\\
&\:\:\:\:-\frac{L^{d-1}\Phi_0^{d-1}(d-1)}{16\pi G_N^{d+1}}\left[\int_{\mathcal{M}_{NHR}}d^2x\sqrt{g^{(2)}}\phi\left(R^{(2)}+\frac{2}{\ell_2^2}\right)+2\int_{\partial \mathcal{M}_{NHR}}dx\sqrt{\gamma^{(1)}}\phi (K^{(1)}-\frac{1}{\ell_2})\right].
\end{split}
\label{eq:nhr_action}
\end{equation}
In the expression we have also included the counterterms (\ref{eq:GH_NHR}) that appeared in the far away region computation below (see Appendix \ref{subsubsec:far_away_region}). The first line of (\ref{eq:nhr_action}) is a consequence of the fact that the potential has a zero at extremality $U(\Phi_0)=0$. Also, the aforementioned line is topological, because of the two-dimensional Gauss-Bonnet theorem\footnote{Recall that the two-dimensional Gauss-Bonet theorem establishes that the Euler character $\chi$ of an euclidean 2d closed surface is given by $4\pi\chi=\int_{\mathcal{M}}\sqrt{g}R+2\int_{\partial \mathcal{M}}K$. For an euclidean disc, $\chi=1$.}. The second line of (\ref{eq:nhr_action}) determines the AdS$_2$ length scale after imposing
\begin{equation}
\frac{U'(\Phi_0)}{(d-1)\Phi_0^{d-2}}=-\frac{2}{\ell_2^2}\:,
\label{eq:first_der_pot}
\end{equation}
From (\ref{eq:first_der_pot}) it follows that
\begin{equation}
\ell_2=\frac{\ell_{d+1}}{\sqrt{d(d-1)}}\:.
\end{equation}
\subsubsection{Far away region}
\label{subsubsec:far_away_region}
The far away region is well approximated by the extremal geometry. It is parametrically defined for: $\tau\in [0, \beta)$, with $\beta=T^{-1}$ being the inverse Hawking temperature of the black hole, and $ r\in [r_{JT} ,r_c)$, with $r_c$ acting as an UV regulator that goes to $r_c \to \infty$, and $r_{JT}$ as the position of the hypersurface connecting the near-horizon region with the far away region. To get the effective JT gravity action (\ref{eq:actiondimRed}), we proceed by evaluating the on-shell action. There are four different contributions that we describe below:
\begin{itemize}
\item The gravitational bulk term is
\begin{equation}
\begin{split}
I_{FAR}^{bulk}&=-\frac{L^{d-1}}{16\pi G_N^{d+1}}\left[\int_{\mathcal{M}_2[FAR]} d^2x\sqrt{g^{(2)}}\left(\Phi^{d-1}R^{(2)}-U(\Phi)\right)\right]\\
&=-\frac{\beta L^{d-1}}{16\pi G_N^{d+1}}\left[(r_{c}^d-r_{JT}^d)\frac{(d-2)}{\ell_{d+1}^2}-\frac{d(3d-4)}{(d-2)}\frac{r_0^{2d-2}}{\ell_{d+1}^2}r_{JT}^{2-d}\right]\\
&=-\frac{\beta L^{d-1}}{16\pi G_N^{d+1}}\left[(d-2)\frac{r_{c}^d}{\ell_{d+1}^2}\right]-\frac{\beta L^{d-1}}{16\pi G_N^{d+1}}\left[\frac{2 r_0^d}{\ell_{2}^2}\right]\phi_b+2\beta M_0\:,
\end{split}
\label{eq:far_bulk}
\end{equation}
where the extremal mass is
\begin{equation}
 M_0=\frac{2(d-1)^2}{(d-2)}\frac{L^{d-1}}{16\pi G_N^{(d+1)}}\frac{r_0^d}{\ell_{d+1}^2}\:.
\end{equation}
Since we are interested in the near-extremal regime, in the last line of (\ref{eq:far_bulk}) we expanded out the position of the boundary surface $r_{JT}$ using (\ref{eq:linaerized_dilaton}), with $r_{JT}=r_0(1+\phi_b)$ for $\phi_b\ll 1$.
\item The Gibbons-Hawking term evaluates to
\begin{equation}
\begin{split}
I_{FAR}^{GH}&=-\frac{L^{d-1}}{16\pi G_N^{(d+1)}}\left[2\int_{\partial \mathcal{M}_2[FAR]}d\tau\:\sqrt{\gamma^{(1)}}\:\Phi^{d-1}K^{(1)}\right]\\
&=-\frac{\beta L^{d-1}}{16\pi G_N^{(d+1)}}\left[d\frac{r_c^d}{\ell_{d+1}^2}\right].
\end{split}
\end{equation}
\item The UV counterterm evaluated at $r_c$, after Weyl rescaling, is
\begin{equation}
\begin{split}
I_{FAR}^{CT}&=-\frac{L^{d-1}}{16\pi G_N^{(d+1)}}\left[-\int_{\partial\mathcal{M}_{2}[FAR]}d\tau\sqrt{\gamma^{(1)}}\left(\frac{2(d-1)}{\ell_{d+1}}\Phi^{\frac{d}{2}}\Phi_0^{\frac{d-2}{2}}\right)\right]\\
&=-\frac{\beta L^{d-1}}{16\pi G_N^{(d+1)}}\left[-2(d-1)\frac{r_c^{d}}{\ell_{d+1}^2}\right]-\beta M_0\:.
\end{split}
\end{equation}
\item The boundary term in the near-horizon region evaluated at $r_{JT}$ from the far away region, with normal vector pointing towards a hypersufarce located at $r=r_c$, 
\begin{equation}
\begin{split}
I_{NHR}^{GH}&=-\frac{L^{d-1}}{16\pi G_N^{d+1}}\left[-2\int_{\partial \mathcal{M}_2[NHR]}d\tau\:\sqrt{\gamma^{(1)}}\:\Phi^{d-1}K^{(1)}\right]\\
&=-\frac{\beta L^{d-1}}{16\pi G_N^{d+1}}\left[-2\frac{r_0^d}{\ell_2^2}\right]\phi_b\:.
\end{split}
\label{eq:GH_NHR}
\end{equation}
Again, we have expanded out $r_{JT}=r_0(1+\phi_b)$ for $\phi_b \ll 1$ as in (\ref{eq:far_bulk}). 
\end{itemize}
Finally, placing all the results together we obtain
\begin{equation}
I_{FAR}^{bulk}+I_{FAR}^{GH}+I_{FAR}^{CT}+I_{NHR}^{GH}=\beta M_0\:.
\label{eq:far_action_final}
\end{equation}
Therefore, the far away region produces a shift of $\beta M_0$ in the on-shell gravitational action, in addition to an extra boundary term in the boundary of the near horizon region at $r=r_{JT}$. Indeed, at linear level in $\phi_b$, the countertems for JT gravity (\ref{eq:nhr_action}) evaluated from the near-horizon region satisfy $I_{JT}^{GH}+I_{NHR}^{GH}=0$.
\subsubsection{Effective action: JT gravity}
\label{subsubsec:JT_app}
Following the line of reasoning of \cite{Maldacena:2016upp}, the total effective action including contributions from the near-horizon region (\ref{eq:far_action_final}) and the far away region (\ref{eq:nhr_action}), can be described by a boundary mode (Schwarzian mode) parameterized by a function $\tau(u)$, with $u$ being the boundary time\footnote{See \cite{Mertens:2022irh,Sarosi:2017ykf} for a review.}. In addition to that, we need to determine the induced boundary conditions on  the metric $g_{\mu \nu}$ and the linearized dilaton $\phi$ at $r=r_{JT}$.

From the solutions to the Einstein-dilaton equations of motion (\ref{eq:sol_EinsDil}), we can read off the induced value for the linearized dilaton at the boundary of the near-horizon region \footnote{The identification $z=\frac{\ell_2^2}{(r-r_0)}$ is motivated at extremality from  the mapping
\begin{equation}
ds^2=\frac{\ell_2^2}{z^2}(d\tau^2+dz^2)\:\:\: \Longleftrightarrow\:\:\:\:ds^2=f(r)d\tau^2+\frac{dr^2}{f(r)}\:\:\:\:\text{with}\:\:\:\:\:\:\:\:\:\: f(r)=\frac{(r-r_0)^2}{\ell_2^2}. \nonumber
\end{equation}.} \cite{Moitra:2019bub}
\begin{equation}
\Phi(r)=r=r_0\left(1+\frac{r-r_0}{r_0}\right)=r_0\left(1+\frac{\ell_2^2}{r_0}\frac{1}{z}\right)\:,
\label{eq:inducedBoundary}
\end{equation}
so that at $r=r_{JT}$, or in terms of a regulator $z=\epsilon$, the linearized dilaton becomes (recall the definition (\ref{eq:linaerized_dilaton}))\begin{equation}
\phi |_{\partial \text{NHR}}=\phi_b=\frac{\phi_r}{\epsilon}\ll 1,\:\:\:\:\:\: \phi_r=\frac{\ell_2^2}{r_0},\:\:\:\:\:\: \epsilon \to 0\:.
\label{eq:dilaton_bdy}
\end{equation}
For the induced metric, we impose Dirichlet boundary conditions
\begin{equation}
ds^2 |_{\partial \text{NHR}}=\frac{du^2}{\epsilon^2}\:.
\label{eq:induced_bdy}
\end{equation}
Further, the effective action, taking into account (\ref{eq:nhr_action}) and (\ref{eq:far_action_final}), is then
\begin{equation}
I^{eff}=\beta M_0-S_0-\tilde{C}\left[\int_{\mathcal{M}_{NHR}}d^2x\sqrt{g^{(2)}}\phi\left(R^{(2)}+\frac{2}{\ell_2^2}\right)+2\int_{\partial \mathcal{M}_{NHR}}dx\sqrt{\gamma^{(1)}}\phi \left(K^{(1)}-\frac{1}{\ell_2}\right)\right]\:,
\end{equation}
with a coupling $\tilde{C}=\frac{(d-1)(r_0L)^{d-1}}{16\pi G_N^{d+1}}$ and boundary conditions (\ref{eq:dilaton_bdy}) and (\ref{eq:induced_bdy}). The equations of motion of the linearized dilaton $\phi$ fixes the Ricci scalar to be constant and negative
\begin{equation}
R^{(2)}=-\frac{2}{\ell_2^2}\:,
\end{equation}
so that the metric is locally AdS$_2$. Therefore, euclidean Poincaré coordinates can be chosen
\begin{equation}
ds^2=\frac{dF^2+dz^2}{z^2}\:.
\end{equation}
The condition (\ref{eq:induced_bdy}) constrains the boundary surface to differ from the euclidean disk boundary located at $z=0$. Indeed, the boundary is located at a surface
\begin{equation}
\text{boundary:  }(F(u),z(u))\:\:,\:\:\:\:\:\:\: z(u)=\epsilon F'(u)\:.
\end{equation}
Finally, the evaluation of the on-shell action gives a pure boundary term, which is governed by the Schwarzian mode $\tau(u)$ living at $r=r_{JT}$.  The result of \cite{Maldacena:2016upp} shows that
\begin{equation}
    I^{eff}_{\text{grav+matter}}=\beta M_0-S_0-C\int_0^{\beta} du \:\text{Sch}\left\{\tan\left(\frac{\pi \tau(u)}{\beta}\right),u\right\},
    \label{eq:actiondimRedAppendix}
\end{equation}
where $\text{Sch}(f(u),u)=\frac{f'''(u)}{f'(u)}-\frac{3}{2}\left(\frac{f''(u)}{f'(u)}\right)^2$ is the Schwarzian derivative of $f(u)$, $\beta=T^{-1}$ is the inverse temperature of the $(d+1)$-dimensional black hole, as well as the extremal mass $M_0$ and the naive extremal entropy $S_0$
\begin{equation}
  M_0=\frac{2(d-1)^2}{(d-2)}\frac{L^{d-1}}{16\pi G_N^{(d+1)}}\frac{r_0^d}{\ell_{d+1}^2}\:\:,\:\:\:\:\:S_0=\frac{r_0^{d-1}}{4 G_N^{(d+1)}}L^{d-1}\:.
\end{equation}
The new effective scale $C$ is 
\begin{equation}
C=2\phi_r\tilde{C}=\frac{1}{2\pi d}\frac{\ell_{d+1}^2 r_0^{d-2}L^{d-1}}{4G_N^{(d+1)}}\:,
\end{equation}
which is related to the thermodynamic mass gap as $M_{\text{gap}}=(2\pi^2 C)^{-1}$.
\section{BTZ ingoing/outgoing solutions}
\label{app:BTZ}
In this Appendix we describe properties of the ingoing/outgoing solutions of the scalar wave equation for the BTZ geometry appearing in Section \ref{subsub:RotatingBTZ}. Some complementary references in the literature are \cite{Birmingham:2001pj,Ghosh:2019rcj,Blake:2019otz}.  

The ingoing solution of the scalar wave equation is
\begin{equation}
G_{\omega, l}^{\text{in}}(z)=\psi_{\omega, l}(z)g_{\omega ,l}(z)\:,
\end{equation}
with
\begin{equation}
\psi_{\omega, l}(z)=z^{-\frac{i}{2}k_+}(1-z)^{\frac{\tilde{\Delta}}{2}}\:_2F_1\left(\frac{\tilde{\Delta}}{2}-\frac{i}{2}(k_{+}+k_{-}),\frac{\tilde{\Delta}}{2}-\frac{i}{2}(k_{+}-k_{-}),1-ik_+,z\right),
\end{equation}
and
\begin{equation}
g_{\omega ,l}(z)=\text{exp}\left[i\left(\omega \int^{\infty}_r \frac{dr'}{f(r')}+l r_+r_- \int^{\infty}_r \frac{dr'}{r'^2f(r')}\right)\right]=\left(\frac{r+r_-}{r-r_-}\right)^{\frac{i k_-}{2}}\left(\frac{r-r_+}{r+r_+}\right)^{\frac{i k_+}{2}}\:.
\end{equation}
The constants appearing in the definitions are
\begin{equation}
k_+=\frac{\omega r_+ + l r_-}{r_+^2-r_-^2},\:\:\:\:\:\:k_-=\frac{\omega r_- + l r_+}{r_+^2-r_-^2},\:\:\:\:\:\:\tilde{\Delta}=2-\Delta=1-\sqrt{1+m^2}\:.
\end{equation}
The outgoing solution can be written in terms of the ingoing one as
\begin{equation}
G_{\omega, l}^{\text{out}}(z)=\psi_{-\omega,- l}(z)g_{\omega ,l}(z)= G_{-\omega, -l}^{\text{in}}(z)g^2_{\omega ,l}(z)\:.
\end{equation}
Of relevant interest is the regime attained when $z\to 1$, where the solutions decay as
\begin{equation}
G_{\omega, l}^{\text{in}}\to a^+_{\omega,l}\left(\frac{1-z}{r_+^2-r_-^2}\right)^{\frac{2-\Delta}{2}}+b^+_{\omega,l}\: \left(\frac{1-z}{r_+^2-r_-^2}\right)^{\frac{\Delta}{2}}+\ldots\:,
\end{equation}
and
\begin{equation}
G_{\omega, l}^{\text{out}}\to a^-_{\omega,l}\left(\frac{1-z}{r_+^2-r_-^2}\right)^{\frac{2-\Delta}{2}}+b^-_{\omega,l}\: \left(\frac{1-z}{r_+^2-r_-^2}\right)^{\frac{\Delta}{2}}+\ldots\:.
\end{equation}
The coefficients $a^{\pm}_{\omega,l},b^{\pm}_{\omega,l}$ turn out to be
\begin{equation}
a^+_{\omega,l}=(r_+^2-r_-^2)^{\frac{2-\Delta}{2}}\frac{\Gamma(1-ik_+)\Gamma(\Delta-1)}{\Gamma\left(\frac{\Delta}{2}-\frac{i}{2}(k_+ +k_-)\right)\Gamma\left(\frac{\Delta}{2}-\frac{i}{2}(k_+-k_-)\right)}\:,
\label{eq:constant_ap}
\end{equation}
\begin{equation}
b^+_{\omega,l}=(r_+^2-r_-^2)^{\frac{\Delta}{2}}\frac{\Gamma(1-ik_+)\Gamma(1-\Delta)}{\Gamma\left(1-\frac{\Delta}{2}-\frac{i}{2}(k_++k_-)\right)\Gamma\left(1-\frac{\Delta}{2}-\frac{i}{2}(k_+-k_-)\right)}\:.
\label{eq:constant_bp}
\end{equation}
Since the outgoing solution satisfies $G_{\omega, l}^{\text{out}}=G_{-\omega, -l}^{\text{in}}$ for $z \to 1$, we also have relations among the different coefficients
\begin{equation}
a^-_{\omega,l}=a^+_{-\omega,-l}\:,\:\:\:\:\:\:\:\:\:\:b^{-}_{\omega,l}=b^{+}_{-\omega,-l\:.}
\label{eq:constans_abm}
\end{equation}
The asymptotic regime towards the AdS$_3$ boundary is given by considering a cutoff hypersurface at $r_c$ with $r_c \gg r_{+},r_{-}$ so that the radial coordinate (\ref{eq:z_r_variables}) gets closer to $z_c \approx 1$ such that
\begin{equation}
z_c=1-\frac{2r_+(r_+-r_-)}{r_c^2}+\ldots\:.
\end{equation}
Hence, towards the conformal AdS$_3$ boundary 
\begin{equation}
G_{\omega, l}^{\text{in}}\big |_{\partial AdS_3} \to a^+_{\omega,l}\: r_c^{-(2-\Delta)}+b^+_{\omega,l}\: r_c^{-\Delta}+\ldots\:,\:\:\:\:\:\:\:\: G_{\omega, l}^{\text{out}}\big |_{\partial AdS_3} \to a^-_{\omega,l} r_c^{-(2-\Delta)}+b^-_{\omega,l} r_c^{-\Delta}+\ldots\:.
\label{eq:AdS3_falloff}
\end{equation}
In the same vein, the near-horizon (NHR) boundary is located at $r_{JT}$ subject to $T\ell_2^2\sim r_+-r_- \ll r_{JT}-r_+ \ll r_{JT}$ \cite{Ghosh:2019rcj}. In that region, the radial coordinate (\ref{eq:z_r_variables}) is also $z_{JT}\approx 1$ with a fall-off given by
\begin{equation}
z_{JT}=1-\frac{2r_+(r_+-r_-)}{r_{JT}^2-r_+^2}+\ldots\:.
\end{equation}
The asymptotic behavior towards the near-horizon region boundary is then
\begin{equation}
G_{\omega, l}^{\text{in}}\big |_{\partial NHR} \to a^+_{\omega,l}\: \left(\frac{r_{JT}-r_+}{r_+-r_-}\right)^{-\frac{(2-\Delta)}{2}}+b^+_{\omega,l}\: \left(\frac{r_{JT}-r_+}{r_+-r_-}\right)^{-\frac{\Delta}{2}}+\ldots\:,
\label{eq:AdS2_in_falloff}
\end{equation}
\begin{equation}
G_{\omega, l}^{\text{out}}\big |_{\partial NHR} \to a^-_{\omega,l}\: \left(\frac{r_{JT}-r_+}{r_+-r_-}\right)^{-\frac{(2-\Delta)}{2}}+b^-_{\omega,l}\: \left(\frac{r_{JT}-r_+}{r_+-r_-}\right)^{-\frac{\Delta}{2}}+\ldots\:.
\label{eq:AdS2_out_falloff}
\end{equation}
The relation between the fall-offs towards $\partial$AdS$_3$ (\ref{eq:AdS3_falloff}) and towards $\partial$NHR (\ref{eq:AdS2_in_falloff}) and (\ref{eq:AdS2_out_falloff}), gives the same linear ratio as obtained in the Appendix B of \cite{Ghosh:2019rcj}, generalizing their result for arbitrary values of momenta $(\omega,l)$.

\bibliography{BibNearExtremalRN}{}
\bibliographystyle{utphys}

\end{document}